\documentclass{article}
\usepackage{amssymb}
\usepackage{latexsym}
\usepackage{amsmath}
\usepackage{multicol}
\usepackage{color,colordvi}
\usepackage{epsf}
\usepackage{graphicx}
\usepackage{multirow}

\newcommand{\paren}[1]{\left(#1\right)}

\newcommand{\PD}[2]{\frac{\partial#1}{\partial#2}}
\newcommand{\PDD}[3]{\frac{\partial^{#1}{#2}}{\partial{#3}^{#1}}}
\newcommand{\lap}[1]{\Delta #1}
\newcommand{\at}[2]{\left. #1 \right|_{#2}}
\newcommand{\grad}[1]{\nabla #1}
\newcommand{\bm}[1]{\boldsymbol{#1}}

\newcommand{\norm}[1]{\left\lVert #1 \right\rVert}

\title{From Three-Dimensional Electrophysiology to the Cable Model: an Asymptotic Study}
\author{Yoichiro Mori\\
University of Minnesota, School of Mathematics\\
206 Church St. SE Minneapolis MN, 55455, U.S.A.\\
{\tt ymori@math.umn.edu.}}
\date{}

\begin{document}
\maketitle

\begin{abstract}
Cellular electrophysiology is often modeled using the cable equations. The cable model 
can only be used when ionic concentration effects and three dimensional geometry
effects are negligible. The {\em Poisson model}, in which the electrostatic potential 
satisfies the Poisson equation and the ionic concentrations satisfy the drift-diffusion 
equation, is a system of equations that can incorporate such effects. The Poisson 
model is unfortunately prohibitively expensive for numerical computation
because of the presence of thin space charge layers at internal membrane boundaries. 
As a computationally efficient and biophysically natural alternative, 
we introduce the {\em electroneutral model} 
in which the Poisson equation is replaced by the electroneutrality condition 
and the presence of the space charge layer is incorporated in boundary conditions 
at the membrane interfaces. 
We use matched asymptotics and numerical computations to show that the 
electroneutral model provides an excellent approximation to the Poisson model. 
Further asymptotic calculations illuminate the relationship of the electroneutral 
or Poisson models with the cable model, and reveal the presence of a hierarchy 
of electrophysiology models.
\end{abstract}

\section{Introduction}\label{intro}

Electrophysiology is the study of the electrical activity of biological tissue \cite{Aidley,Hille}.
Because of its importance in many physiological processes and
its quantitative nature, it has been a favorite subject in biophysics and mathematical physiology.
Traditional mathematical models of cellular electrical activity are based on the famous work 
of Hodgkin and Huxley \cite{HH}, and may be collectively termed cable models.
These models are based upon an ohmic current continuity 
relation on a branched one dimensional electrical cable \cite{Koch,KS}.
The derivation of the cable model is based on several important assumptions \cite{Koch}:
\begin{itemize}
\item A one dimensional picture, or more generally, a one dimensional tree representation
of cell geometry is adequate. Geometrical details that are lost in making this simplified 
description have negligible effect on electrophysiology.
\item The extracellular space can be reduced to a single isopotential electrical 
compartment.
\item Ionic concentrations are effectively constant in space and time within 
each cell separately and in the extracellular space. The diffusive current 
that may be induced by concentration gradients or the changes in equilibrium 
potential are negligible.
\end{itemize}
Such assumptions are justified in many instances, for example in the isolated 
neuronal axon \cite{HH},  where the cable model has been extremely successful in explaining 
the physiology and in making quantitative predictions -- a triumph counted 
among the greatest successes of mathematics in biology.
There may, however, be many cases in which any or all of the above assumptions are 
violated especially in the central nervous system and cardiac tissue, as suggested by the 
complex microhistological structure they exhibit \cite{Kandel,Bers}.
One line of work that addressed this difficulty was that of Qian and Sejnowski \cite{Qian-Sej1}.
Their work addresses the last of the above difficulties, but retains the one-dimensional 
character of the cable model.

In \cite{MPJ1}, we presented a three-dimensional model of cellular electrical 
activity which addresses all of the above limitations of the cable model. 
This model consists of a system of partial differential equations
to be satisfied by the ionic concentrations and the electrostatic potential.
In this paper, we introduce a slight modification of this model,
which we call the {\em electroneutral model}. 

The first goal of this paper is to demonstrate the validity of the 
electroneutral model by comparing this with the {\em Poisson model} 
\cite{Leonetti1}.
In the Poisson model, 
the ionic concentration dynamics is governed by the drift-diffusion 
equations and the electrostatic potential satisfies the Poisson equation.
Non-dimensionalization reveals the presence of multiple temporal scales 
and of a thin boundary layer at the membrane interfaces in which 
electric charge accumulates (Debye layer)\cite{Rubinstein}. 
This boundary layer necessitates the use of a fine spatiotemporal mesh 
in numerical simulations making such computations prohibitively expensive.
We introduce the electroneutral model as an alternative to the Poisson model, 
in which the Poisson equation is replaced by the electroneutrality condition.
The model does not resolve the dynamics within the thin boundary layers
and instead incorporates the effect of these layers by modifying 
the boundary conditions at the membranes. The boundary layers are 
incorporated as charge densities of zero thickness at the membrane, 
a picture that is better aligned with the biophysical 
view of the membrane being a capacitor within a conducting medium.
This obviates the necessity 
for high spatiotemporal resolution in computations, making the 
electroneutral model far more amenable to numerical simulation. 
Using matched asymptotics, we show that the electroneutral model
provides an approximation to the Poisson model. 
We present computational studies in the final section to demonstrate 
that the electroneutral model does indeed provide 
an excellent approximation to the 
Poisson Model for biophysically realistic parameter values.

The second goal of this paper is to clarify the relationship between the 
Poisson and electroneutral models to cable models.
If we are to claim that the Poisson or electroneutral models are
a generalization of the cable model, we would like to know 
under what conditions these models can be reduced to the cable model.
Continuing with the asymptotic calculations above, we show that the 
cable model can be obtained as an asymptotic limit under assumptions. 
We shall see that 
there is a hierarchy of electrophysiology models, 
the Poisson or electroneutral models being the most detailed, 
and the traditional cable model being the simplest.

\section{Poisson Model}
\begin{figure}
\begin{center}
\includegraphics[width=0.75\textwidth]{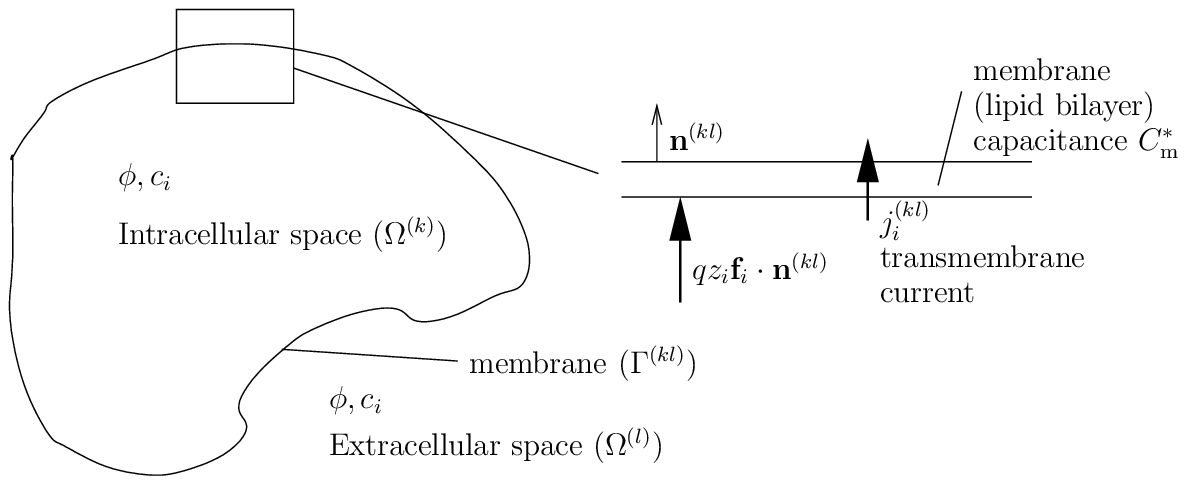}
\end{center}
\caption{The variables $\phi,c_i$ are defined in the regions $\Omega^{(k)}$ and 
 $\Omega^{(l)}$, which we have identified as intracellular and extracellular 
regions in the above. The membrane acts primarily as a capacitor, but possesses
ionic channels through which transmembrane current can flow.}\label{bcfig}
\end{figure}

We first present the Poisson model, which is essentially equivalent 
to the model proposed in \cite{Leonetti1}.
We consider biological tissue to be a three-dimensional space partitioned 
into the intracellular and extracellular spaces by the membrane.
Let the biological tissue of interest be divided into subregions 
$\Omega^{(k)}$, indexed by $k$. We denote the membrane separating 
the regions $\Omega^{(k)}$ and $\Omega^{(l)}$ by $\Gamma^{(kl)}$
(Figure \ref{bcfig}).

In $\Omega^{(k)}$, the equations to be satisfied by the 
ionic concentration $c_i$ and the electrostatic potential $\phi$ are the following.
\begin{align}
\PD{c_i}{t} &= -\nabla \cdot {\mathbf{f}_i} &\text{(ion conservation)} \label{orig1}\\
\mathbf{f}_i &= -D_i\paren{\grad{c_i} +\frac{qz_ic_i}{k_BT}\grad{\phi}} 
&\text{(drift-diffusion flux)}\label{orig2}\\
\lap{\phi} &= -\frac{1}{\epsilon}\paren{\rho_0 + \sum_{i=1}^{N} qz_ic_i} 
&\text{(Poisson equation)} \label{orig3}
\end{align}
Here, $\mathbf{f}_i$ denotes the flux of the $i$-th ion.  $\mathbf{f}_i$ 
is expressed as a sum of two terms, the diffusion term and the drift term.  
$D_i$ is the diffusion coefficient of the $i$-th ion, $qz_i$ is the amount of 
charge on the $i$-th ion, where $q$ is the elementary charge, i.e., the charge 
on a proton.  $qz_iD_i/(k_BT)$ is the mobility of the ion species (Einstein relation) 
where $k_B$ is the Boltzmann constant, and $T$ 
the absolute temperature.  
Fixed background charge density (if any) is given by $\rho_0$, and $\epsilon$ 
is the dielectric constant of the electrolyte solution.
We note that the above system of equations has been used extensively in semiconductor
device modeling \cite{WVR,Jerome_semiconductor} and ionic channel modeling \cite{Nonner,KS,Peskin_neuron}.

Biological membranes consists largely of a lipid bilayer that acts as 
a capacitor impermeable to ions.
In this lipid bilayer are embedded ionic channels and transporters
through which certain ionic species may pass.
With this picture in mind, we write down the boundary conditions 
for the above system to be satisfied at both faces of the membrane.

Consider the boundary condition for the Poisson equation.  
The value of the electrostatic potential and the 
normal component of the electric displacement vector 
$\mathbf{D} = \epsilon \mathbf{E}$,
where $\epsilon$ is the dielectric constant and $\mathbf{E}$ is the electric field,
should be continuous 
at the interface between the cell membrane and the electrolyte solution.  
Therefore, at this interface,
\begin{align}
\phi^{(\text{mem})}&=\phi^{(k)}\\
\epsilon_\text{m}\PD{\phi^{(\text{mem})}}{\mathbf{n}^{(kl)}}&=\epsilon\PD{{\phi^{(k)}}}{\mathbf{n}^{(kl)}} 
\label{bcpoisson}
\end{align}
where $\phi^{(\text{mem})}$ is the electrostatic potential within the membrane, 
$\epsilon_m$ the dielectric constant of the 
cell membrane, and $\mathbf{n}^{(kl)}$ the unit normal at the membrane-electrolyte interface 
pointing from $\Omega^{(k)}$ into the membrane.

We note that (\ref{bcpoisson}) is not satisfied 
at the mouths of ion channels. If ion channels mouths
do not occupy a significant amount of membrane area, the above boundary condition 
may be deemed reasonable. Fortunately, ion channels are sparsely distributed 
even at their peak documented densities \cite{Koch}.

The membrane thickness $d_\text{m}(\sim 10 \text{nm})$ is small compared 
to the curvature radius of the membrane and the typical length scale of the system.
This implies that $\phi^{\text{mem}}$ varies
linearly as one traverses the membrane from $\Omega^{(k)}$ to $\Omega^{(l)}$.  
Thus,
\begin{equation}
\PD{\phi^{(\text{mem})}}{\mathbf{n}}=\frac{\phi^{(k)}-\phi^{(l)}}{d_\text{m}}.  
\end{equation}
We obtain the following boundary condition,
\begin{equation}
C_\text{m}^{*}\phi^{(kl)}=\epsilon\PD{\phi^{(k)}}{\mathbf{n}^{(kl)}}  \label{origBC1}
\end{equation}
where $\phi^{(kl)}=\phi^{(k)}-\phi^{(l)}$, $C_\text{m}^{*}= \frac{\epsilon_\text{m}}{d_\text{m}}$ and 
${\mathbf{n}^{(kl)}}$ is the unit normal on the membrane pointing from $\Omega^{(k)}$ to $\Omega^{(l)}$.
$C_\text{m}^{*}$ may be considered the intrinsic capacitance of the membrane, which is to be 
distinguished from the effective membrane capacitance 
$C_\text{m}$ to appear later. The jump in the electrostatic potential $\phi^{(kl)}$
is termed the {\em membrane potential} and is one of the primary biophysical quantities of interest.

The boundary conditions for the drift diffusion equations are simple:
\begin{equation}
qz_i\mathbf{f}_i\cdot \mathbf{n}^{(kl)}=j^{(kl)}_i  \label{origBC2}
\end{equation}
where $j^{(kl)}_i$ are ion channel currents carried by the $i$-th species of ion.
We note that $j^{(kl)}_i=-j^{(lk)}_i$.
These currents can in general be functions of the ionic concentrations of arbitrary species 
on either side of the membrane, the membrane potential $\phi^{(kl)}$ and 
gating variables which describe the internal states of a given ionic channel \cite{KS,MPJ1}.

We shall refer to equations (\ref{orig1})-(\ref{orig3}) supplemented 
with boundary conditions (\ref{origBC1}) and (\ref{origBC2}), 
as the {\em Poisson model}.

\section{Non-Dimensionalization and Multiple Spatiotemporal Scales}\label{nondimmult}
We non-dimensionalize the Poisson model.
We first rescale the ionic concentrations $c_i$ 
and the electrostatic potential $\phi$ as follows.
\begin{align}
\phi&=\frac{k_BT}{q}\Phi, & c_i&=c_0C_i & &\\
\rho_0&=qc_0\tilde{\rho_0},& \mathbf{f}_i&=c_0\tilde{\mathbf{f}_i},& j_i&=\gamma qc_0\tilde{j_i}
\end{align}
where $c_0\approx 100 mmol/l$ is the characteristic concentration and 
$\gamma qc_0$ is the characteristic magnitude of the 
transmembrane current per unit area. $k_BT/q\approx 25mV$ is the natural 
unit for the membrane potential. 
The constant $\gamma$ has units of velocity$=$length$/$time and its 
typical physiological range is:
\begin{equation}
\gamma \approx 10^{-5}\sim 10^{-3} \mu\text{m}/\text{msec}\label{gam}
\end{equation}

We determine a typical length scale of the system.
We take equation (\ref{origBC2}) and integrate over the membrane surface $\partial\Omega^{(k)}$.
\begin{equation}
\begin{split}
\int_{\partial\Omega^{(k)}} \gamma \tilde{j_i} dA
&=\int_{\partial\Omega^{(k)}}z_i\tilde{\mathbf{f}_i}\cdot \mathbf{n}^{(kl)}dA
=\int_{\Omega^{(k)}}z_i\nabla\cdot\tilde{\mathbf{f}_i}dV\\
&=-\int_{\Omega^{(k)}} z_i\nabla\cdot D_iC_i(\nabla \mu_i)dV
\end{split}
\end{equation}
where we have used dimensionless variables for ionic concentration and the electrostatic potential.
In the above, $dV$ and $dA$ denote volume and surface integrals respectively and 
$\mu_i$ is the chemical potential $\ln C_i+z_i\Phi$.  
We have used the divergence theorem in the second equality 
and the flux expression (\ref{orig2}) in the third.
Let $L_0$ be the typical length over which the flux and the chemical potential vary.  
Balancing the order of magnitude of the surface and volume integrals above,
\begin{equation}
\gamma |\partial\Omega^{(k)}|=\frac{D_i|\Omega^{(k)}|}{L_0^2}.
\end{equation}
where $|\partial \Omega^{(k)}|$ is the surface area of the region $\Omega^{(k)}$ and 
$|\Omega^{(k)}|$ is the volume of $\Omega^{(k)}$.
We therefore set:
\begin{equation}
L_0=\sqrt{\frac{lD_0}{\gamma}}, \text{ where } \quad l=\frac{|\Omega^{(k)}|}{|\partial\Omega^{(k)}|}.
\label{defL0}
\end{equation}
The constant $D_0\approx 1\mu m/msec^2$ is the typical diffusion coefficient for ions.
The quantity $l$ is a measure of the volume per unit surface area, 
and is a representative length scale of the distance between membranes.  
For a cylindrical axon, $l$ corresponds
roughly to the diameter of the axon.
As we shall see in Section \ref{cbl}, $L_0$ is what is termed the electrotonic length in cable theory.
Notice that $L_0$ is proportional to $\sqrt{l}$. This is in agreement with the 
observation in cable theory that the electrotonic length scales with the square root the 
diameter of a cylindrical cable \cite{KS}.

Given $L_0$, we can define a typical time scale $T_0$ as
$T_0=L_0^2/D_0=l/\gamma$. This expression tells us that $T_0$
is equivalently the time scale in which the dimensionless ionic concentration
experiences changes of $\mathcal{O}(1)$.
We shall call $T_0$ the {\em diffusion time scale} or the 
{\em slow diffusion time scale}. 

Using $L_0$ and $T_0$ as the representative spatiotemporal scales,
we introduce the following dimensionless variables.
\begin{align}
\mathbf{x}&=L_0\mathbf{X}, & t&=T_0\tau_D,
& D_i&=D_0\tilde{D_i}\\
\tilde{\mathbf{f}_i}&=\frac{D_0}{L_0}\mathbf{F}_i & \alpha&=\frac{l}{L_0}
\end{align}
We can now write the Poisson model (\ref{orig1})-(\ref{orig3}) and (\ref{origBC1})-(\ref{origBC2})
in dimensionless form:
\begin{align}
\PD{C_i}{\tau_D}&=-\nabla_{\mathbf{X}} \cdot \mathbf{F}_i \label{nd1}\\
\mathbf{F}_i&=-\tilde{D_i}(\nabla_{\mathbf{X}} C_i + z_iC_i\nabla_{\mathbf{X}} \Phi)\label{nd2}\\
\beta^2\Delta_{\mathbf{X}}\Phi&=-(\tilde{\rho_0}+\sum_{i=1}^N z_iC_i)\label{nd3}
\end{align}
The boundary conditions are, 
\begin{align}
\theta^* \Phi^{(kl)}&=\beta \PD{\Phi}{\mathbf{n}^{(kl)}}\label{ndBC1}\\
z_i\mathbf{F}_i \cdot \mathbf{n}^{(kl)}&=\alpha\tilde{j_i}\label{ndBC2}
\end{align}
Note that $\alpha$ is the dimensionless magnitude of the transmembrane currents 
as well as the dimensionless volume to surface ratio.
We have introduced the dimensionless parameters $\beta$ and 
$\theta^*$. The parameter $\beta$ is the ratio between the {\em Debye length} $r_d$ \cite{Rubinstein}
and $L_0$:
\begin{equation}
\beta= \frac{r_d}{L_0}, \quad r_d\equiv \sqrt{\frac{\epsilon k_B T}{q^2c_0}}
\end{equation}
The Debye length is typically $r_d\approx 1nm$, and is considerably smaller 
than the typical length scale $L_0$. 
The parameter $\theta^*$ is defined as follows:
\begin{equation}
\theta^*=\frac{C_\text{m}^*}{\epsilon/r_d}=\frac{C_\text{m}^*k_BT/q}{qc_0r_d}\approx 10^{-2} \label{thetadef}
\end{equation}
We have, thus, three constants $\beta, \alpha$ and $\theta^*$ that characterize the 
system.

Given typical values of $l$ and $\gamma$, 
we can find typical physiological values of the parameters $\beta$ and $\alpha$
(the magnitude of $\theta^*$ is given in (\ref{thetadef}).).
Recall that $l$ is the (dimensional) volume to surface ratio, and thus, roughly measures the separation 
distance of membranes.  Values typical in the central nervous system can range from 
$100\text{nm}$ to $10\mu\text{m}$. Combining this with the radius of $\gamma$ (\ref{gam}),
we obtain the following physiological ranges for the above parameters.
\begin{align}
L_0&=\sqrt{\frac{lD_0}{\gamma}}=10\mu\text{m}\sim 1\text{mm} & & \\  
\beta&=r_d\sqrt{\frac{\gamma}{lD_0}}=10^{-6}\sim 10^{-4}, & 
\alpha&=\sqrt{\frac{l\gamma}{D_0}}=10^{-3}\sim 10^{-1} \label{alph}
\end{align}
We note that while the magnitude of $\beta$ and $\alpha$ depend on the geometry ($l$)
and electrophysiological properties ($\gamma$) of the physiological system 
under consideration, $\theta^*$ defined in (\ref{thetadef})
is a constant that varies little between physiological systems. 

We shall exploit the smallness of the parameter $\beta$ to reduce the Poisson model.
Note that $\beta^2$ multiplies the Laplacian in (\ref{nd2}). 
By formally taking $\beta \rightarrow 0$ in (\ref{nd2}), we see 
that the electroneutrality condition:
\begin{equation}
\tilde{\rho_0}+\sum_{i=1}^N z_iC_i=0
\end{equation}
should be approximately satisfied in the bulk of the region of interest.
The electroneutrality condition above is in general not compatible with the 
mixed (Robin) boundary condition of (\ref{ndBC1}), and thus, we have 
a singular perturbation problem which gives rise to a boundary layer at the membrane. 
Given that $\beta^2$ multiplies a second spatial derivative in (\ref{nd2}),
a layer of $\mathcal{O}(\beta)$ develops at the membrane, where electric 
charge may accumulate. In dimensional terms, this layer has thickness 
$r_d \sim 1nm$ near the membrane. We shall refer to this layer as the {\em space charge layer}
or {\em Debye layer}. This is a layer that we have no need to resolve as long 
as we are interested in electrophysiology at the cellular or subcellular level 
and not at the molecular level. The biophysical 
equivalent of this layer in the cable model is the charge associated with the 
membrane modeled as a capacitor, and accordingly, has no spatial extent. 
Starting with Section \ref{matchedasymp}, we shall
perform a matched asymptotic calculation that addresses the presence of this layer.

We can now interpret the dimensionless parameter $\theta^*$
in (\ref{thetadef}) as follows.  The constant $k_B T / q$ is the typical magnitude 
of the membrane potential, whereas $q c_0r_d$ is a natural unit of
surface charge density since $r_d$ gives the surface charge thickness.
Thus, $q c_0r_d / (k_B T / q)$ is a
natural unit of capacitance per unit area.  The constant $\theta^*$
expresses the membrane capacitance per unit area in these
natural units.

Before we can perform asymptotics on the model, we would like to identify 
other spatiotemporal scales that the Poisson model possesses.
Differentiate both sides of equation (\ref{nd3}) in $\tau_D$ and 
take the integral over $\Omega^{(k)}$. The left hand side yields:
\begin{equation}
\begin{split}
\int_{\Omega^{(k)}}\PD{}{\tau_D}(\beta^2\Delta \Phi) dV
&=\int_{\partial\Omega^{(k)}}\PD{}{\tau_D}\paren{\beta^2\PD{\Phi}{\mathbf{n}^{kl}}}dA\\
&=\int_{\partial\Omega^{(k)}}\beta\theta^*\PD{\Phi^{(kl)}}{\tau_D}dA. \label{ndpleft}
\end{split}
\end{equation}
We used the boundary condition (\ref{ndBC1}) in the second inequality.
The right hand side yields:
\begin{equation}
-\int_{\Omega^{(k)}}\PD{}{\tau_D}\paren{\tilde{\rho_0}+\sum_{i=1}^N z_iC_i}dV
=\int_{\partial\Omega^{(k)}}\alpha \sum_{i=1}^N\tilde{j_i}^{(kl)}dA. \label{ndpright}
\end{equation}
where we have used (\ref{nd1}), (\ref{ndBC1}) and the divergence theorem.
The above says that the change in total charge within $\Omega^{(k)}$ comes from transmembrane currents.
Balancing the quantities in (\ref{ndpleft}) and (\ref{ndpright}), we see that
the membrane potential and hence the electrostatic potential can vary 
on the time scale of $\beta\frac{\theta^*}{\alpha}T_0$. 
It is an interesting coincidence that $\theta^*$ and $\alpha$  
are roughly of the same order of magnitude, as can be seen from (\ref{thetadef}) and (\ref{alph}).  
Thus, this time scale is roughly equal to $\beta T_0$, which we shall call the 
{\em membrane potential time scale}. Given the smallness of $\beta$, the membrane potential time scale
is considerably smaller than the slow diffusion time scale $T_0$.  
 We shall see in Section \ref{cbl} that the membrane potential time scale $\beta T_0$ corresponds to the 
``diffusion'' time scale of the membrane potential in the traditional cable model.

There is yet another time scale, which corresponds to charge relaxation:
\begin{equation}
\begin{split}
\PD{}{\tau_D}\paren{\tilde{\rho_0}+\sum_{i=1}^N z_iC_i}
&=\sum_{i=1}^N (z_i\nabla\cdot\tilde{D_i}\nabla C_i 
+ z_i^2\nabla (\tilde{D_i}C_i) \cdot \nabla \Phi + z_i^2C_i\Delta \Phi)\\
&=-\paren{\sum_{i=1}^N z_i^2C_i}\frac{1}{\beta^2}\paren{\tilde{\rho_0}+\sum_{i=1}^N z_iC_i} + \text{other terms} 
\end{split}
\end{equation}
where we have used the Poisson equation (\ref{nd3}) in the last equality to replace $\Delta \Phi$.  
We see that charge density decays exponentially with a time constant of $\beta^2T_0=r_d^2/D_0=1\text{nsec}$.
We can infer that this time scale is only important where the electrolyte solution may 
deviate significantly from electroneutrality, i.e., within the space charge layer.

We thus see that there are three time scales present in the Poisson model, 
$T_0$, $\beta T_0$ and $\beta^2T_0$.
We list the physiological values for these time scales.
\begin{equation}
T_0=10^{-1}\sim 10^3\text{ sec}, \quad \beta T_0=10^{-2}\sim 1\text{ msec}, \quad \beta^2 T_0=1\text{ nsec}
\end{equation}
The time scale of greatest interest is the $\beta T_0$ time scale, in which the
membrane potential varies. This is also roughly equal to the time scale in which the 
most rapid physiological processes take place, such as
channel gating, chemical neurotransmission and calcium concentration changes \cite{Hille}.  
We shall thus focus our attention on this time scale and rescale 
the time variable $\tau_D$ to a newly rescaled time variable $\tau_V\equiv\tau_D/\beta$.
We write $C_i,\Phi$ as functions of $\tau_V$ rather than $\tau_D$.
Equation (\ref{nd1}) is rescaled to:
\begin{align}
\PD{C_i}{\tau_V}&=-\beta \nabla_{\mathbf{X}} \cdot \mathbf{F}_i \label{ndm1}
\end{align} 
The $\beta^2T_0$ time scale and the space charge layer within which this time 
scale is relevant are spatiotemporal details that we have no need to resolve. 
The $T_0$ time scale is important with regard to long term changes in ionic concentrations.
We shall make some brief remarks about this time scale in the final section.

An overarching goal is to computationally investigate the three dimensional 
electrical activity of complex physiological systems. 
A great difficulty with the Poisson model is that one inevitably needs to resolve 
spatiotemporal scales associated with the space charge layer in a numerical simulation,
making such computations prohibitively expensive.
It would therefore be computationally desirable to have a model 
that resolves the membrane potential time scale but 
does not resolve the Debye spatiotemporal scales.

\section{Electroneutral Model}\label{electroneutral_model}
We propose the following as a computationally efficient alternative to the 
Poisson Model:
\begin{align}
0&=\PD{C_i}{\tau_V}+\beta\nabla_{\mathbf{X}} \cdot \mathbf{F}_i \label{WEN1}\\
\mathbf{F}_i&=-\tilde{D_i}(\nabla_{\mathbf{X}} C_i + z_iC_i\nabla_{\mathbf{X}} \Phi)\label{WEN2}\\
0&=\tilde{\rho_0}+\sum_{i=1}^N z_iC_i\label{WEN3}\\
z_i\mathbf{F}_{i}\cdot \mathbf{n}^{(kl)}&=\PD{\sigma_i^{(k)}}{\tau_V}+\alpha\tilde{j}_i
\label{WEN4}
\end{align}
The Poisson equation in the Poisson model has been replaced 
by the electroneutrality condition (\ref{WEN3}).
Since this is an algebraic condition, it does 
not require a boundary condition at the membrane.
The boundary conditions for the drift-diffusion equations (\ref{WEN1}) and (\ref{WEN2}) are 
given by (\ref{WEN4}). In comparison to (\ref{ndBC2}), we have an additional term:
\begin{equation}
\PD{\sigma_i}{\tau_V}.
\end{equation}
$\sigma_i$ is the amount of electric charge at the membrane face contributed by the $i$-th species of ion.
In the electroneutral model, the electric charge within the Debye layer is represented 
as a surface charge density of zero thickness. In this picture, the amount of ionic current 
$qz_i\mathbf{f}_i\cdot \mathbf{n}$ either contributes to the change in surface charge density 
$\sigma_i$ or flows across the membrane through ion channels. 
This picture is better aligned with the biophysical view of the membrane in the cable model,
in which the membrane is a capacitor within an ohmic medium.
One important advantage of the boundary 
condition (\ref{WEN4}) compared with (\ref{ndBC2}) is that the parameter values in (\ref{WEN4})
are directly observable experimentally. 
Since the Debye layers are too thin to be explored experimentally,
the parameter values in (\ref{ndBC2}) can only be inferred, as argued in detail in \cite{MPJ1}.
 
The surface charge contributions $\sigma_i$ must be related to the dynamic 
variables $C_i$ and/or $\Phi^{(kl)}$ to close the system of equations.
First we let
\begin{equation}
\sum_{i=1}^N \sigma_i\equiv \sigma =\theta \Phi^{(kl)}.
\end{equation}
This relation says that the total amount of surface charge $\sigma$ is 
linearly proportional to the membrane potential $\Phi^{(kl)}$, where $\theta$
is the effective dimensionless membrane capacitance. Note that $\theta$ is different from 
$\theta^*$, the {\em intrinsic} membrane capacitance, 
used in (\ref{ndBC1}). The lipid bilayer sandwiched by the two boundary layers considered as 
a whole gives rise to a capacitor with the effective capacitance $\theta$. 
This is the capacitance that is measured experimentally, given that it is impossible to 
to distinguish the contributions to the capacitance from the Debye layers and the lipid bilayer.
The relation between these two quantities will 
be clarified in Appendix \ref{app1}.
Now, define $\lambda_i$ as the fraction of the total charge $\sigma$ that is 
contributed by the $i$-th species of ion:
\begin{equation}
\sigma_i=\lambda_i\sigma.\label{deflambdai1}
\end{equation}
We let $\lambda_i$ evolve according to the following:
\begin{equation}
\PD{\lambda_i}{\tau_V}=\frac{\tilde{\lambda}_i-\lambda_i}{\beta}, \quad \tilde{\lambda}_i=\frac{z_i^2C_i}{\sum_{i'=1}^{N} z_{i'}^2C_{i'}}
\label{WEN5}
\end{equation}
The charge fraction $\lambda_i$ relaxes to $\tilde{\lambda}_i$ in the charge relaxation time scale.
The specific form of $\tilde{\lambda}_i$ was derived in \cite{MPJ1}, 
but is also given in Appendix \ref{app1}.
Note that:
\begin{equation}
\beta \PD{}{t}\paren{\sum_{i=1}^N \lambda_i} 
= \sum_{i=1}^N(\tilde{\lambda}_i-\lambda_i)=1-\paren{\sum_{i=1}^N \lambda_i}\label{lambdasum1}
\end{equation}
and therefore, $\sum_{i=1}^N\lambda_i\equiv 1$ provided that $\sum_{i=1}^N\lambda_i= 1$ 
at the initial time, as required by the definition of $\lambda_i$ as the charge fraction. 
In \cite{MPJ1}, $\tilde{\lambda}_i$ was used in place of $\lambda_i$ in (\ref{deflambdai1}), 
in which case the charge fraction relaxation equation in (\ref{WEN5}) is not needed. 
This original system, however, leads to ill-posed behavior which we 
examine in Appendix \ref{app2}. 

We shall call the system (\ref{WEN1})-(\ref{WEN4}) and (\ref{WEN5}) the {\em electroneutral model}.
There is no longer a space charge layer to be resolved at the membrane, since the presence of the 
surface charge has been taken care of in the boundary condition (\ref{WEN4}). 
The charge relaxation time scale
only appears in a simple ODE (\ref{WEN5}), and does not pose serious difficulties in the 
construction of a numerical scheme \cite{morithesis}. We propose the electroneutral model 
as a computationally tractable model 
that addresses the shortcomings of the cable model pointed out in Section \ref{intro}.

An important difference between the electroneutral model and the Poisson model is what the state variables are.
In the Poisson model, specifying the ionic concentrations at every point in space is enough to describe the 
state of the system. The electrostatic potential can be found from the ionic concentration profile by solving 
the Poisson equation (\ref{nd3}) with the boundary conditions (\ref{ndBC1}). 
The difficulty, though, is that we must specify the ionic concentrations 
{\em up to the boundary to within the space charge layer}. 
The electroneutral model, on the other hand, does {\em not} require the 
ionic concentration profiles in the space charge layer. 
The spatiotemporal details of the space charge layer are represented by the 
the membrane potential $\Phi^{(kl)}$ and the charge fractions $\lambda_i$.
The state variables 
for the electroneutral model thus include the ionic concentration profile 
as well as the membrane potential $\Phi^{(kl)}$
and the membrane charge fractions $\lambda_i$. This means in particular that 
we need to specify the values of these quantities as initial conditions.

In the electroneutral model we have 
ion conservation in the following sense:
\begin{equation}
\PD{}{\tau_V}\paren{\int_{\Omega^{(k)}}z_iC_i dV+\int_{\Gamma^{(kl)}} \beta \theta \lambda_i^{(k)}\Phi^{(kl)}dA}
=-\int_{\Gamma^{(kl)}} \beta\alpha \tilde{j}_idA.\label{consvion}
\end{equation}
This equation says that for each ionic species 
the change in the sum of the ionic content of the region $\Omega^{(k)}$
and of the space charge layer is equal to the transmembrane current that flows out of this region. 
This is an important property not only from a physical point of view, but also from a practical point of 
view if we are to perform long-time calculations of ionic concentration dynamics.

The natural question that arises is whether the electroneutral model is in any way an 
approximation to the Poisson model. We investigate this question using both asymptotic 
and numerical computations. Beginning with the next section, we present a matched asymptotic study 
to show that the electroneutral model gives an approximation to 
the Poisson model. In Section \ref{comp}, we shall computationally investigate 
how well the electroneutral model approximates the Poisson model.

\section{Matched Asymptotics}\label{matchedasymp}
We recall the Poisson Model:
\begin{align}
\PD{C_i}{\tau_V}&=-\beta \nabla_{\mathbf{X}} \cdot \mathbf{F}_i\\
\mathbf{F}_i&=-\tilde{D_i}(\nabla_{\mathbf{X}} C_i + z_iC_i\nabla_{\mathbf{X}} \Phi)\\
\beta^2\Delta_{\mathbf{X}}\Phi&=-\paren{\tilde{\rho_0}+\sum_{i=1}^N z_iC_i}
\end{align}
Recall from (\ref{ndm1}) that we rescaled time to $\tau_V$ to capture 
the dynamics in the membrane potential time scale. We now perform matched 
asymptotics on the above to clarify the relation between the electroneutral 
and Poisson models.

As noted earlier, a boundary layer of thickness $\mathcal{O}(\beta)$ develops at the membrane
when $\beta \ll 1$.
We therefore introduce an {\em inner layer} of thickness $\mathcal{O}(\beta)$ at the membrane.
We shall continue to use the terms space charge layer or Debye layer to denote this layer.

\begin{figure}
\begin{center}
\includegraphics[width=0.75\textwidth]{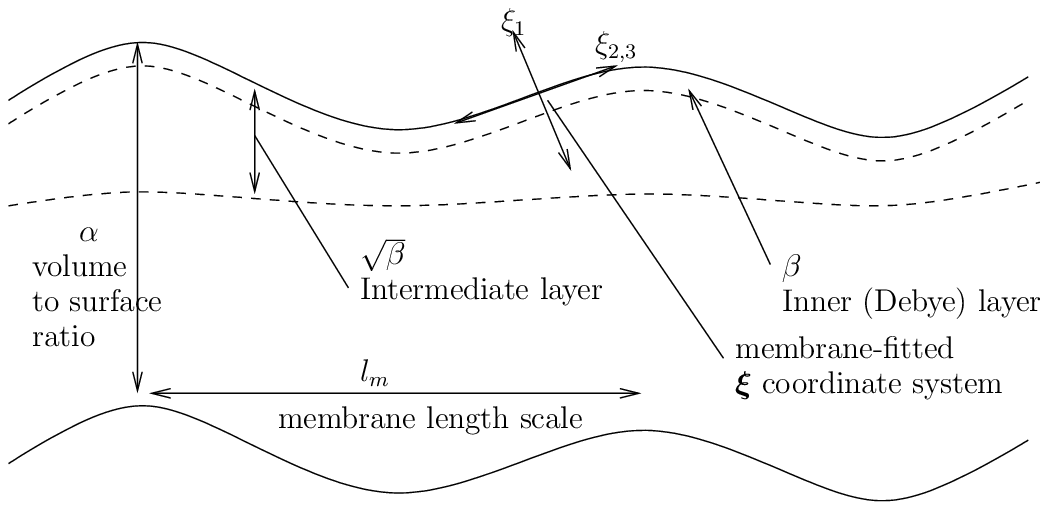}
\end{center}
\caption{A schematic of the relevant spatial scales used in the asymptotic calculations.
The solid lines denote the membrane and the dotted lines are the interfaces between 
the boundary layers. The inner-most layer has width $\beta$, the intermediate layer $\sqrt{\beta}$.
The typical membrane separation is $\alpha$ and the typical length scale associated 
with the membrane is $l_m$. $\bm{\xi}$ is the membrane-fitted coordinate used in 
the matched asymptotics calculations. }\label{layerfig}
\end{figure}

We need to introduce another spatial scale of order $\mathcal{O}(\sqrt{\beta})$
at the membrane. This need
arises as the result of introducing a newly rescaled time variable $\tau_V$. 
The spatial scale of order $\mathcal{O}(\sqrt{\beta})$ corresponds 
to the length over which ions can diffuse in the membrane potential time scale, $\beta T_0$.
Formally, the necessity for this layer can be seen by noting that $\beta$ multiplies 
a second order spatial derivative in (\ref{ndm1}) since
$\mathbf{F}_i$ is itself written in terms of spatial derivatives (c.f. \ref{nd2}).
We shall refer to this layer as the {\em intermediate layer} or the {\em fast diffusion layer}.
It is interesting to note that the presence of such layers have been  
postulated to account for K$^+$ ion accumulation in the extracellular space of 
the squid giant axon \cite{Frankenhaeuser-Hodgkin}.
We thus have three regions to consider in the asymptotic calculations to follow:
the inner and intermediate layers located adjacent to the membrane, 
and the region away from the membrane,
which we shall call the {\em outer layer}. We perform two 
matching procedures, at the inner-intermediate layer interface and at the intermediate-outer layer
interface. We have summarized the relevant spatial scales in Figure \ref{layerfig}.

The above discussion prompts us to expand the physical variables in powers of $\sqrt{\beta}$ 
instead of $\beta$:
\begin{align}
C_i(\mathbf{X},\tau_V)&
=C_i^0(\mathbf{X},\tau_V)+\sqrt{\beta}C_i^1(\mathbf{X},\tau_V)+\beta C_i^2(\mathbf{X},\tau_V)\cdots \label{cform}\\
\Phi(\mathbf{X},\tau_V)&
=\Phi^0(\mathbf{X},\tau_V)+\sqrt{\beta}\Phi^1(\mathbf{X},\tau_V)+\beta \Phi^2(\mathbf{X},\tau_V)\cdots \label{phiform}
\end{align}
The other two parameters of the system, $\alpha$ and $\theta^*$ are also small 
(c.f. (\ref{alph}),(\ref{thetadef})), but 
we shall treat them as being $\mathcal{O}(1)$ with respect to $\beta$.  We note that 
$\beta$ is typically a few orders of magnitude smaller than $\alpha$ or $\theta^*$.
The smallness of $\alpha$ and $\theta^*$ will be later exploited, 
in sections \ref{cbl} and \ref{inintmatch} respectively. 

In performing matched asymptotics at the membrane, we introduce a coordinate system at the
membrane $\bm{\xi}=(\xi_1,\xi_2,\xi_3)$, where the $\xi_1$ axis is taken to be perpendicular 
to the membrane, while $\xi_2$ and $\xi_3$ are curvilinear coordinates that 
run ``parallel'' to the membrane.  
The $\xi_1$ axis will be rescaled to yield coordinates in the 
intermediate layer $\bm{\xi}^a$ such that $\xi_1=\sqrt{\beta}\xi_1^a$ and in the inner layer
$\bm{\xi}^b$ such that $\xi_1=\beta\xi_1^b$. 

We must now ask how we are to rescale $\xi_2$ and $\xi_3$. 
There are at least two spatial scales that are relevant:
$\rho_\kappa$ the dimensionless curvature radius of the membrane 
and $l_j$ the dimensionless length scale on which one may see $\mathcal{O}(1)$ changes in ion channel current 
density. Let $l_m$ be the smaller of the two spatial scales  $l_j$ and $\rho_\kappa$. We shall 
call $l_m$ the \textit{membrane length scale}. The question raised at the beginning of 
this paragraph can be answered by comparing the relative magnitude of this length scale
to the $\mathcal{O}(\sqrt{\beta})$ length scale.

If $l_m$ is considerably 
larger than $\sqrt{\beta}$, there is no need to rescale $\xi_2$ and $\xi_3$. 
If $l_m$ is order $\mathcal{O}(\sqrt{\beta})$, we must 
scale $\xi_2, \xi_3$ to $\xi_2=\sqrt{\beta}\xi_2^{a,b}, \xi_3=\sqrt{\beta}\xi_3^{a,b}$
so that the curvature correction and the ionic fluxes parallel to the membrane are 
$\mathcal{O}(1)$ quantities when measured in the intermediate layer coordinate $\bm{\xi}^a$.  
We shall mainly be concerned with the case $l_m > \sqrt{\beta}$ but 
we shall quote results of calculations when $l_m\sim \sqrt{\beta}$.

We point out that there could be situations in which $l_m$ is small only along a certain coordinate 
direction.  For example, if we take a cylindrical axon with diameter $\mathcal{O}(\sqrt{\beta})$, 
and take $\xi_2$ to be the angular coordinate, and $\xi_3$ to be the axial coordinate, 
the curvature radius along the $\xi_2$ coordinate is $\mathcal{O}(\sqrt{\beta})$ whereas the curvature 
radius along the $\xi_3$ coordinate is large (curvature is negligible).  
In such cases (and if the cylindrical axon is endowed with near uniform 
ion channel density so that $l_j$ is large), we need only rescale $\xi_2$ but not $\xi_3$.
We shall not deal with such cases, since such an analysis will follow along similar lines
as the case in which $l_m\sim \sqrt{\beta}$.

\section{Inner-Intermediate Matching}\label{inintmatch}

We first consider inner-intermediate matching when $l_m>\sqrt{\beta}$.

Consider the membrane surface facing $\Omega^{(k)}$.
We now introduce a coordinate system $\bm{\xi}$ 
so that the $\xi_1$ coordinate direction is perpendicular to the membrane.
We let $\xi_1=0$ coincide with the membrane face, 
and let the positive $\xi_1$ axis point into the region $\Omega^{(k)}$. 
For simplicity, we shall assume that the membrane is flat, i.e., that it has no curvature.
Therefore, we can take the coordinate system to $\bm{\xi}$ to be orthonormal.
When $l_m>\sqrt{\beta}$, it turns out that curvature corrections produce only higher order terms
that we can ignore.

In the inner layer, we rescale $\bm{\xi}$ as:
\begin{equation}
\bm{\xi}^b=(\xi_1^b,\xi_2^b,\xi_3^b), \quad \xi_1=\beta \xi_1^b, \quad \xi_2=\xi_2^b, \quad \xi_3=\xi_3^b.
\end{equation}
The equations are:
\begin{align}
\beta\PD{C_i^b}{\tau_V}
&=-\paren{\PD{F_{i1}^b}{\xi_1^b}+\beta^2\paren{\PD{F_{i2}^b}{\xi_2^b}+\PD{F_{i3}^b}{\xi_3^b}}}\label{ndmi1}\\
F_{ip}^b&=-\tilde{D_i}\paren{\PD{C_i^b}{\xi_p^b} + z_iC_i^b\PD{\Phi^b}{\xi_p^b}}, \quad p=1,2,3 \label{ndmi2}\\
\PDD{2}{\Phi^b}{\xi_1^b}+\beta^2\paren{\PDD{2}{\Phi^b}{\xi_2^b}+\PDD{2}{\Phi^b}{\xi_3^b}}
&=-\paren{\tilde{\rho_0}+\sum_{i=1}^N z_iC_i^b}.\label{ndmi3}
\end{align}
Since the inner layer is adjacent to the membrane, 
we must supplement the above with boundary conditions, suitably rescaled:
\begin{align}
\theta^* (\at{\Phi}{\xi_1^b=0}-\Phi^{(l)})&=-\at{\PD{\Phi^b}{\xi_1^b}}{\xi_1^b=0}\label{ndmBC1}\\
-\at{z_iF_{i1}^b}{\xi_1^b=0}&=\beta\alpha\tilde{j}_i.\label{ndmBC2}
\end{align}
We shall make the simplifying 
assumption that the transmembrane ionic current
densities $j_i$ are given functions of position (on the membrane)
and time instead of being functions of $C_i, \Phi^{(kl)}$ and the gating variables.

In the intermediate layer we rescale $\bm{\xi}$ as:
\begin{equation}
\bm{\xi}^a=(\xi_1^a,\xi_2^a,\xi_3^a),\quad \xi_1=\sqrt{\beta} \xi_1^a, \quad \xi_2=\xi_2^a, \quad \xi_3=\xi_3^a.
\end{equation}
The equations are:
\begin{align}
\PD{C_i^a}{\tau_V}
&=-\paren{\PD{F_{i1}^a}{\xi_1^a}+\beta\paren{\PD{F_{i2}^a}{\xi_2^a}+\PD{F_{i3}^a}{\xi_3^a}}}\label{ndma1}\\
F_{ip}&=-\tilde{D_i}\paren{\PD{C_i^a}{\xi_p^a} + z_iC_i^a\PD{\Phi^a}{\xi_p^a}}, \quad p=1,2,3 \label{ndma2}\\
\beta \PDD{2}{\Phi^a}{\xi_1^a}+\beta^2\paren{\PDD{2}{\Phi^a}{\xi_2^a}+\PDD{2}{\Phi^a}{\xi_3^a}}
&=-\paren{\tilde{\rho_0}+\sum_{i=1}^N z_iC_i^a}.\label{ndma3}
\end{align}

Substitute (\ref{cform}) and (\ref{phiform}) 
in the inner layer equations (\ref{ndmi1})-(\ref{ndmi3}),
and collect like terms in order $\beta$. The expansions of $C_i$ and $\Phi$ in $\sqrt{\beta}$ 
induce expansions of $\mathbf{F}_i$ in terms of $\sqrt{\beta}$. We shall denote 
the $\mathcal{O}(\sqrt{\beta}^k)$ term as $\mathbf{F}_i^k$. For example, 
\begin{align}
F_{i1}^{b0}&=-\tilde{D_i}\paren{\PD{C_i^{b0}}{\xi_1^b}+z_iC_i^{b0}\PD{\Phi^{b0}}{\xi_1^b}}\\
F_{i1}^{b1}&=-\tilde{D_i}\paren{\PD{C_i^{b1}}{\xi_1^b}+z_iC_i^{b1}\PD{\Phi^{b0}}{\xi_1^b}+z_iC_i^{b0}\PD{\Phi^{b1}}{\xi_1^b}}
\end{align}
By applying the same procedure to the equations (\ref{ndma1})-(\ref{ndma3}), 
we may obtain analogous expressions in the intermediate layer.

We derive matching conditions at the inner-intermediate layer interface 
in terms of the ionic fluxes.
Note from (\ref{ndmi1}) and (\ref{ndmBC2}) that:
\begin{align}
\PD{F_{i1}^{b0}}{\xi_1^b}&=0, \quad \at{F_{i1}^{b0}}{\xi_1^b=0}=0\\
\PD{F_{i1}^{b1}}{\xi_1^b}&=0, \quad \at{F_{i1}^{b1}}{\xi_1^b=0}=0
\end{align}
From this, we find that
\begin{equation}
F_{i1}^{b0}=F_{i1}^{b1}\equiv 0\label{F=0}
\end{equation}
within the inner layer. 

Now, consider the $p=1$ component of (\ref{ndmi2}) and (\ref{ndma2}), 
$F_{i1}^b$ and $F_{i1}^a$.
We introduce a matching coordinate system $\xi^\eta$ in between the inner and intermediate 
layers such that,
\begin{equation}
\xi_1^a=\eta(\beta) \xi^\eta, \quad 
\lim_{\beta\rightarrow 0}\frac{\sqrt{\beta}}{\eta}=\lim_{\beta\rightarrow 0}\eta=0
\end{equation}
Applying Kaplun's matching condition \cite{holmes,keener-app}
to  $F_{i1}^b$ and $F_{i1}^a$, we obtain:
\begin{align}
\lim_{\beta\rightarrow 0}&F_{i1}^{b0}\paren{\frac{\eta}{\sqrt{\beta}}\xi^\eta}=0\label{matchF0}\\
\lim_{\beta\rightarrow 0}&\paren{
\frac{1}{\sqrt{\beta}}F_{i1}^{b0}\paren{\frac{\eta}{\sqrt{\beta}}\xi^\eta}
+F_{i1}^{b1}\paren{\frac{\eta}{\sqrt{\beta}}\xi^\eta}-F_{i1}^{a0}(\eta \xi^\eta)}=0\label{matchF1}\\
\nonumber
\lim_{\beta\rightarrow 0}&\left(
\frac{1}{\beta}F_{i1}^{b0}\paren{\frac{\eta}{\sqrt{\beta}}\xi^\eta}
+\frac{1}{\sqrt{\beta}}F_{i1}^{b1}\paren{\frac{\eta}{\sqrt{\beta}}\xi^\eta}
+F_{i1}^{b2}\paren{\frac{\eta}{\sqrt{\beta}}\xi^\eta}\right.\\
&\quad \left.-\frac{1}{\sqrt{\beta}}F_{i1}^{a0}(\eta \xi^\eta)
-F_{i1}^{a1}(\eta \xi^\eta)\right)=0\label{matchF2}
\end{align}
Condition (\ref{matchF0}) is automatically satisfied by (\ref{F=0}).
Condition (\ref{matchF1}), taken together with (\ref{F=0}), yields: 
\begin{equation}
\lim_{\beta\rightarrow 0}F_{i1}^{a0}(\eta \xi^\eta)=\at{F_{i1}^{a0}}{\xi_1^a=0}=0.\label{matchF11}
\end{equation} 
We thus have the matching condition for the leading order ionic flux in the intermediate layer.
The last matching condition (\ref{matchF2}), combined with (\ref{F=0}), yields the following.
\begin{equation}
\lim_{\beta\rightarrow 0}\left(
F_{i1}^{b2}\paren{\frac{\eta}{\sqrt{\beta}}\xi^\eta}
-\frac{1}{\sqrt{\beta}}F_{i1}^{a0}(\eta \xi^\eta)
-F_{i1}^{a1}(\eta \xi^\eta)\right)=0. \label{matchF}
\end{equation}

To evaluate (\ref{matchF}), we 
need to calculate $C_i$ and $\Phi$ to leading order in the inner layer.
From (\ref{F=0}), (\ref{ndmi3}) and (\ref{ndmBC1}) we see
that the leading order terms satisfy the following one dimensional 
boundary value problem in $\xi_1^a$ in the inner layer:
\begin{align}
0&=\PD{C_i^{b0}}{\xi_1^b}+z_iC_i^{b0}\PD{\Phi^{b0}}{\xi_1^b}\label{in2}\\
\PDD{2}{\Phi^{b0}}{\xi_1^b}&=-(\tilde{\rho_0}+\sum_{i=1}^N z_iC_i^{b0})\label{in4}\\
\theta^* \paren{\Phi^{b0}(\xi_1^b=0)-\Phi^{(l),b0}(\xi_1^{b,(l)}=0)}&=-\at{\PD{\Phi^{b0}}{\xi_1^b}}{\xi_1^b=0}\label{in5}\\
C_i^{b0}(\xi_1^b=\infty)&=C_i^{a0}(\xi_1^a=0)\label{mat1}\\
\Phi^{b0}(\xi_1^b=\infty)&=\Phi^{a0}(\xi_1^a=0).\label{mat2}
\end{align}
The last two conditions come from matching conditions at the inner\--inter\-mediate layer interface.
Here, $\xi_1^{b,(l)}$ refers to the inner layer coordinate system on the $\Omega^{(l)}$
side of the membrane $\Gamma^{(kl)}$(note that we are now working on the $\Omega^{(k)}$ side). 
Equations (\ref{in2}), (\ref{in4}) with boundary conditions (\ref{in5})-(\ref{mat2})
can be solved explicitly under the approximation that $\theta^*$ is small, a reasonable 
approximation since $\theta^*\approx 10^{-2}$ (cf. (\ref{thetadef})).
We quote the results below, and relegate the calculations to Appendix \ref{app1}.
\begin{align}
\Phi^{b0}(\xi_1^b)&=\Phi^{a0}(0)-\frac{\tilde{\sigma}}{\Gamma}\exp(-\Gamma \xi_1^b)\label{Phiin}\\
C_i^{b0}(\xi_1^b)&=C_i^{a0}(0)\paren{1+\frac{z_i\tilde{\sigma}}{\Gamma}\exp(-\Gamma \xi_1^b)}\label{Ciin}\\
\Gamma^2&=\sum^{N}_{i=1}z_i^2C_i^{a0}(0), \quad \Gamma>0 \label{gamma}\\
\tilde{\sigma}_i&\equiv
\frac{z_i^2C_i^{a0}(0)}{\Gamma^2}\tilde{\sigma}\equiv \tilde{\lambda}_i\tilde{\sigma}\label{tildesigmaexp}\\
\tilde{\sigma}&=\theta\Phi^{(kl),a0},\quad 
\frac{1}{\theta}=\frac{1}{\theta^*}+\frac{1}{\Gamma^{(k)}}+\frac{1}{\Gamma^{(l)}}
\end{align}
In the above, $\tilde{\sigma}$ denotes the total charge in the Debye layer, and 
$\tilde{\sigma}_i$ is the charge contributed by the $i$-th species of ion. 
Thus, $\tilde{\sigma}_i/\tilde{\sigma}=\tilde{\lambda}_i$ is the charge fraction
contributed by the $i$-th species of ion. Note by design that 
$\sum_{i=1}^N\tilde{\lambda}_i=1$.
The variable $\theta$ is the dimensionless {\em effective} membrane capacitance
to be distinguished from the dimensionless {\em intrinsic} membrane capacitance $\theta^*$.
We refer the reader to Appendix \ref{app1} for further elaboration.


Now, consider (\ref{ndmi1}) and (\ref{ndmBC2}) at the first non-trivial order:
\begin{equation}
\PD{C_i^{b0}}{\tau_V}
=-\PD{F_{i1}^{b2}}{\xi_1^b}, \quad -\at{z_iF_{i1}^{b2}}{\xi_1^b=0}=\alpha\tilde{j}_i \label{matstartlarge}
\end{equation}
Since our goal is to evaluate (\ref{matchF}), 
we would like to obtain an expression for $F_{i1}^{b2}$. We integrate the above 
in $\xi_1^b$ to obtain:
\begin{equation}
\begin{split}
-z_iF_{i1}^{b2}&=\alpha\tilde{j}_i+
z_i\int_0^{\xi_1^b}\PD{C_i^{b0}}{\tau_V}d\xi_1^b\\
&=\alpha\tilde{j}_i+z_i\PD{C_i^{a0}(0)}{\tau_V}\xi_1^b
+\int_0^{\xi_1^b}\PD{}{\tau_V}\frac{z_iC_i^{a0}(0)\tilde{\sigma}}{\Gamma}\exp(-\Gamma \xi_1^b)d\xi_1^b\\
&\equiv\alpha\tilde{j}_i+z_i \PD{C_i^{a0}(0)}{\tau_V}\xi_1^b+I^\text{charge} \label{chargeterm}
\end{split}
\end{equation}
where we used (\ref{Ciin}) for $C_i^{b0}$.
We can finally consider condition (\ref{matchF}). 
We would like (\ref{matchF}) be satisfied regardless of the value of $\xi^\eta$.
For the $F_{i2}^{b2}$ term, taking $\beta\rightarrow 0$ in (\ref{matchF}) amounts 
to studying the behavior of (\ref{chargeterm}) in the limit $\xi_1^b\rightarrow\infty$. 
Take $\xi_1^b\rightarrow\infty$ in $I^\text{charge}$.
\begin{equation}
\lim_{\xi_1^b\rightarrow \infty} z_iI^{\text{charge}}=\PD{\tilde{\sigma}_i}{\tau_V}.
\end{equation}
where we have used (\ref{tildesigmaexp}).
We thus conclude using (\ref{chargeterm}) and the above that:
\begin{equation}
-z_iF_{i1}^{b2}= \paren{\PD{\tilde{\sigma}_i}{\tau_V}+\alpha\tilde{j}_i}
+z_i\PD{C_i^{a0}(0)}{\tau_V}\xi_1^b +\mathcal{O}(\exp(-\Gamma \xi_1^b))\label{mF}
\end{equation}
Note that $F_{i1}^{b2}$ thus consists of a constant and a linear component in $\xi_1^b$
as well as a residual term that decays exponentially.
We now expand the intermediate layer expressions of (\ref{matchF}) at $\xi^\eta=0$.
\begin{equation}
\begin{split}
\frac{1}{\sqrt{\beta}}F_{i1}^{a0}(\eta \xi^\eta)+F_{i1}^{a1}(\eta \xi^\eta)
&=\frac{1}{\sqrt{\beta}}F_{i1}^{a0}(0)
+\frac{\eta \xi^\eta}{\sqrt{\beta}}\PD{F_{i1}^{a0}(0)}{\xi_1^a}+F_{i1}^{a1}(0)+\cdots\\
&=\frac{\eta \xi^\eta}{\sqrt{\beta}}\PD{F_{i1}^{a0}(0)}{\xi_1^a}+F_{i1}^{a1}(0)+\cdots
\end{split}
\end{equation}
where we have used (\ref{matchF11}) to eliminate $F_{i1}^{a0}(0)$.
Substituting the above as well as (\ref{mF}) into (\ref{matchF}), 
\begin{equation}
\begin{split}
\lim_{\beta\rightarrow 0}&\paren{F_{i1}^{a1}(0)+\frac{1}{z_i}\paren{\PD{\tilde{\sigma}_i}{\tau_V}+\alpha\tilde{j}_i}}\\
&+\frac{\eta \xi^\eta}{\sqrt{\beta}}\paren{\PD{F_{i1}^{a0}(0)}{\xi_1^a}+\PD{C_i^{a0}(0)}{\tau_V}}
+\mathcal{O}(\exp(-\Gamma \xi_1^b))+\cdots\\
&=0
\end{split}
\end{equation}
The necessary conditions for the above to be satisfied are:
\begin{align}
-z_iF_{i1}^{a1}(0)&=\PD{\tilde{\sigma}_i}{\tau_V}+\alpha\tilde{j}_i \label{matlarge}\\
\PD{C_i^{a0}(0)}{\tau_V}&=-\PD{F_{i1}^{a0}(0)}{\xi_1^a}\label{matlarge2}
\end{align}
The second expression (\ref{matlarge2}) is automatically satisfied as can be seen by taking (\ref{ndma2})
to leading order. 
Equation (\ref{matlarge}) together with (\ref{matchF11}) 
are the matching condition we set out to obtain.

When $l_m\sim \sqrt{\beta}$, as discussed at the end of the 
previous section, we must rescale the coordinates 
so that $\xi_p=\sqrt{\beta}\xi_p^{a,b}, \quad p=2,3$.
We can obtain the matching conditions for this case 
in a manner similar to the $l_m > \sqrt{\beta}$ case, although
the calculations are more involved.
The matching conditions corresponding to (\ref{matlarge}) and (\ref{matlarge2}) are
respectively \cite{morithesis}:
\begin{align}
-z_iF_{i1}^{a1}(0)
&=\PD{\tilde{\sigma}_i}{\tau_V}+\alpha\tilde{j}_i
-\nabla_{S^a}\cdot\paren{\tilde{D_i}\tilde{\sigma}_i\nabla_{S^a} (\ln C_i^{a0}(0)+z_i\Phi^{a0}(0))}
\label{matsmall}\\
\PD{C_i^{a0}(0)}{\tau_V}&=-\paren{\PD{F_{i1}^{a0}(0)}{\xi_1^a}+\nabla_{S^a}\cdot\mathbf{F}_{iS^a}^{a0}(0)}
\label{matsmall2}
\end{align}
Here, the operators $\nabla_{S^a}$ and $\nabla_{S^a} \cdot$ denote respectively the gradient 
and divergence operators on the membrane, where the length is measured in terms of $\sqrt{\beta}L_0$.

Compared with (\ref{matlarge}), equation (\ref{matsmall}) has 
an additional membrane drift diffusion term.
The surface gradient of the chemical potential 
potential $\mu_i=\ln C_i^{a0}+z_i\Phi^{a0}$, scaled by the diffusion coefficient,
gives the drift velocity of $\tilde{\sigma_i}$ along the membrane.

We shall henceforth limit our attention to the case $l_m>\beta$.

\section{Electroneutral Model as Approximation to Poisson Model}\label{ENPo}
We now examine the relationship between the Poisson and electroneutral models. 
Consider two pairs of ionic concentrations and electrostatic potential
$C_i^\text{EN},\Phi^\text{EN}$ and $C_i^\text{Po},\Phi^\text{Po}$,
which evolve according to the electroneutral and Poisson models respectively. 
We postulate an expansion of $C_i^\text{EN},C_i^\text{Po}$ and 
$\Phi^\text{EN},\Phi^\text{Po}$ in $\sqrt{\beta}$ of the form (\ref{cform}) and (\ref{phiform})
respectively and see if the electroneutral and Poisson models produce 
the same leading order equations. 

First consider the intermediate layer. We write equations 
(\ref{WEN1})-(\ref{WEN4}) and (\ref{WEN5}) of the electroneutral model 
in the $\bm{\xi}^a$ coordinate and write out the leading order equations.
The $\mathcal{O}(1)$ equations are:
\begin{align}
\PD{C_i^{a0,\text{EN}}}{\tau_V}&=-\PD{F_{i1}^{a0,\text{EN}}}{\xi_1^a}\label{EN1start}\\
F_{i1}^{a0,\text{EN}}&=-\tilde{D_i}\paren{\PD{C_i^{a0,\text{EN}}}{\xi_1^a} + z_iC_i^{a0,\text{EN}}\PD{\Phi^{a0,\text{EN}}}{\xi_1^a}}\\
0&=\tilde{\rho_0}+\sum_{i=1}^N z_iC_i^{a0,\text{EN}}\\
F_{i1}^{a0,\text{EN}}(\xi_1^a=0)&=0\label{EN1end}
\end{align}
The $\mathcal{O}(\sqrt{\beta})$ equations are:
\begin{align}
\PD{C_i^{a1,\text{EN}}}{\tau_V}&=-\PD{F_{i1}^{a1,\text{EN}}}{\xi_1^a}\label{EN1/2start}\\
F_{i1}^{a1,\text{EN}}&=-\tilde{D_i}\paren{\PD{C_i^{a1,\text{EN}}}{\xi_1^a} + z_iC_i^{a1,\text{EN}}\PD{\Phi^{a0,\text{EN}}}{\xi_1^a}
+z_iC_i^{a0,\text{EN}}\PD{\Phi^{a1,\text{EN}}}{\xi_1^a}}\\
0&=\tilde{\rho_0}+\sum_{i=1}^N z_iC_i^{a1,\text{EN}}\\
F_{i1}^{a1,\text{EN}}(\xi_1^a=0)&=\theta\PD{\lambda_i\Phi^{(kl)a0,\text{EN}}}{\tau_V}+\alpha\tilde{j}_i\label{EN1/2end-1}\\
\PD{\lambda_i}{\tau_V}&=\frac{\tilde{\lambda}_i^\text{EN}-\lambda_i}{\beta}, 
\tilde{\lambda}_i^\text{EN}=\frac{z_i^2C_i^{a0,\text{EN}}}{\sum_{i'=1}^{N} z_{i'}^2C_{i'}^{a0,\text{EN}}}\label{EN1/2end}
\end{align}
The same procedure on the Poisson model yields the following.
The $\mathcal{O}(1)$ equations are:
\begin{align}
\PD{C_i^{a0,\text{Po}}}{\tau_V}&=-\PD{F_{i1}^{a0,\text{Po}}}{\xi_1^a}\label{Po1start}\\
F_{i1}^{a0,\text{Po}}&=-\tilde{D_i}\paren{\PD{C_i^{a0,\text{Po}}}{\xi_1^a} + z_iC_i^{a0,\text{Po}}\PD{\Phi^{a0,\text{Po}}}{\xi_1^a}}\\
0&=\tilde{\rho_0}+\sum_{i=1}^N z_iC_i^{a0,\text{Po}}\\
F_{i1}^{a0,\text{Po}}(\xi_1^a=0)&=0 \label{Po1end}
\end{align}
Equation (\ref{Po1end}) comes from the matching condition (\ref{matchF11}).
The $\mathcal{O}(\sqrt{\beta})$ equations are:
\begin{align}
\PD{C_i^{a1,\text{Po}}}{\tau_V}&=-\PD{F_{i1}^{a1,\text{Po}}}{\xi_1^a}\label{Po1/2start}\\
F_{i1}^{a1,\text{Po}}&=-\tilde{D_i}\paren{\PD{C_i^{a1,\text{Po}}}{\xi_1^a} + z_iC_i^{a1,\text{Po}}\PD{\Phi^{a0,\text{Po}}}{\xi_1^a}
+z_iC_i^{a0,\text{Po}}\PD{\Phi^{a1,\text{Po}}}{\xi_1^a}}\\
0&=\tilde{\rho_0}+\sum_{i=1}^N z_iC_i^{a1,\text{Po}}\\
F_{i1}^{a1,\text{Po}}(\xi_1^a=0)&=\theta\PD{\tilde{\lambda}^\text{Po}_i\Phi^{(kl)a0,\text{Po}}}{\tau_V}+\alpha\tilde{j}_i\label{Po1/2end-1}\\
\tilde{\lambda}_i^\text{Po}&=\frac{z_i^2C_i^{a0,\text{Po}}}{\sum_{i'=1}^{N} z_{i'}^2C_{i'}^{a0,\text{Po}}}\label{Po1/2end}
\end{align}
where equations (\ref{Po1/2end-1}) and (\ref{Po1/2end}) come from the matching condition (\ref{matlarge}).

We see that (\ref{EN1start})-(\ref{EN1end}), (\ref{EN1/2start})-(\ref{EN1/2end-1}) are identical to 
(\ref{Po1start})-(\ref{Po1end}), (\ref{Po1/2start})-(\ref{Po1/2end-1}), except for the difference between 
$\lambda_i$ and $\tilde{\lambda}_i$ in equation (\ref{EN1/2end-1}) and (\ref{Po1/2end-1}). In Appendix \ref{app2}
we show that in fact (Eq. (\ref{diffdeltlambda})):
\begin{equation}
\PD{\lambda_i^\text{EN}}{\tau_V}=\PD{\tilde{\lambda}_i^\text{EN}}{\tau_V}+\mathcal{O}(\beta)
\end{equation}
Therefore, $\lambda_i^\text{EN}$ may be replaced by $\tilde{\lambda}_i^\text{EN}$ without 
affecting the order of the approximation.
This shows that $C_i^\text{EN},\Phi^\text{EN}$ and $C_i^\text{Po},\Phi^\text{Po}$ satisfy identical 
equations in the intermediate layer to order $\mathcal{O}(\sqrt{\beta})$.

The same procedure in the outer layer shows that the two models agree up to 
equations of order $\mathcal{O}(\beta^{3/2})$. 
We thus see that the electroneutral model formally approximates the 
Poisson model in the intermediate layer and outer layers, where the biophysical 
processes of interest take place.
In Section \ref{comp}, we shall show computationally that 
the electroneutral model indeed provides an excellent approximation to the Poisson model.

\section{Equations in the Outer Layer}

We continue with the asymptotic calculations with the goal of obtaining the cable model
under certain conditions to be set forth below.

\subsection{3D Cable Model}\label{3dcable}
We now consider intermediate-outer matching.
We first turn to the 
equations satisfied in the outer layer, 
which can be obtained by substituting (\ref{phiform}) and (\ref{cform}) 
into (\ref{ndm1}), (\ref{nd2}) and (\ref{nd3}).
\begin{align}
\PD{C_i^0}{\tau_V}&=0, & \PD{C_i^1}{\tau_V}&=0 \label{out1}\\
\PD{C_i^2}{\tau_V}&=-\nabla_\mathbf{X} \cdot \mathbf{F}_i^0\label{out2} & &\\
0&=\tilde{\rho_0}+\sum_{i=1}^N z_iC_i^0 &  0&=\sum_{i=1}^N z_iC_i^2\quad \label{out4}
\end{align}
Equation (\ref{out1}) tells us that $C_i$ to leading order does not change in the $\tau_V$ time 
variable. We still need to 
know the evolution of $\Phi^0$. This can be obtained by summing (\ref{out2}) in $i$ and 
and using (\ref{out4}) to conclude:
\begin{equation}
\nabla\cdot\paren{\sum_{i=1}^N z_i\mathbf{F}_i^0}=0 \label{eqphiout}
\end{equation}
This is the equation satisfied by $\Phi^0$ in the outer layer. 
In order to obtain the boundary condition for this 
equation, all we need is $(\sum_{i=1}^N z_i\mathbf{F}_i^0)\cdot \mathbf{n}^{(kl)}$.

Let $\mathcal{J}=\sum_{i=1}^N z_i\mathbf{F}_i$. We shall use the usual subscripts and superscripts
on $\mathcal{J}$ to denote terms of the expansion of $\mathcal{J}$ in $\beta$ in the different layers, induced by 
the expansion of $\mathbf{F}_i$. We find from (\ref{ndma1}) and (\ref{ndma3}) that:
\begin{align}
\sum_{i=1}^N z_iC_i^{a0}&=0, & \sum_{i=1}^N z_iC_i^{a1}&=0 \label{int2}\\
\PD{C_i^{a0}}{\tau_V}&=-\PD{F_{i1}^{a0}}{\xi_1^a}, &  \PD{C_i^{a1}}{\tau_V}&=-\PD{F_{i1}^{a1}}{\xi_1^a}\label{int3}
\end{align}
Using the above relations we see that:
\begin{equation}
\PD{\mathcal{J}_1^{a0}}{\xi_1^a}=\PD{\mathcal{J}_1^{a1}}{\xi_1^a}=0\label{constJ}
\end{equation}
The value of $\mathcal{J}_1^{a0}$ and $\mathcal{J}_1^{a1}$ at $\xi_1^a=0$
can be computed from (\ref{matchF11}) and (\ref{matlarge}), 
and we see from (\ref{constJ}) that:
\begin{equation}
\mathcal{J}_1^{a0}=0, \quad -\mathcal{J}_1^{a1}=\theta\PD{\Phi^{(kl),a0}}{\tau_V}+\sum_{i=1}^N\alpha\tilde{j}_i
\end{equation}
where we have used $\sum_{i=1}^N \tilde{\lambda}_i=1$.
Following the same matching procedure as for the inner-intermediate layer matching, we conclude:
\begin{equation}
\theta\PD{\Phi^{(kl),a0}}{\tau_V}+\alpha\sum_{i=1}^N \tilde{j}_i=-\mathcal{J}^0 \label{matchintout}
\end{equation}
We can now use the above as the boundary condition for (\ref{eqphiout}) and
explicitly write down the equations satisfied in the outer layer. 
\begin{align}
\nabla\cdot(A\grad\Phi^0+\grad{B})&=0 \label{3dc1}\\
-(A\grad\Phi^0+\grad{B})\cdot\mathbf{n}^{(kl)}&=\theta\PD{\Phi^{(kl),0}}{\tau_V}+\alpha I_\text{ion}\label{3dc2}\\
A=\sum_{i=1}^N z_i^2C_i^0, &\quad B=\sum_{i=1}^N z_iC_i^0, \quad I_\text{ion}=\sum_{i=1}^N\tilde{j}_i\label{3dc3}
\end{align}
Note here that $A$ and $B$ are functions of $\mathbf{X}$ only, and do not depend 
on time, since $C_i^0$ does not change in the $\tau_V$ time scale. 

There is one difficulty here that needs to be pointed out. Equation (\ref{matchintout}) and (\ref{3dc2}) 
are not exactly the same. In (\ref{matchintout}), $\Phi^{(kl)}$ is evaluated just outside the inner layer, 
whereas in (\ref{3dc2}), $\Phi^{(kl)}$ is evaluated just outside the intermediate layer. There is a similar 
concern for the transmembrane current terms $\tilde{j}_i$ if they are functions of $C_i$ or $\Phi^{(kl)}$.

From (\ref{int2}) and (\ref{matchF11}), and the fact that $C_i^{a0}$ and $\Phi^{a0}$ must match to leading 
order at $\xi_1^a=\infty$ to the outer layer solution, we see that $C_i^{a0}$ and $\Phi^{a0}$ decay to a 
uniform state after an initial transient (note $C_i^{a0}$ decays to a constant where as $\Phi^{a0}$ decays
to a time-varying uniform state, whose value is equal to $\Phi^0(\xi_1^b=0)$).  Therefore, after 
an initial transient, the discrepancy between $\Phi^{(kl),0}$, $\Phi^{(kl),a0}$ and $C_i^{a0}$, $C_i^0$
will decay to $0$.

This model is valid to leading order outside 
the intermediate layer of thickness $\mathcal{O}(\sqrt{\beta})$. 
We shall call this the {\em 3D-cable model}.

The the 3D-cable model may be derived very easily from the electroneutral model.
Consider equations (\ref{WEN1})-(\ref{WEN4}) of the electroneutral model.
We can take the time derivative of the electroneutrality condition (\ref{WEN3}) and 
substitute (\ref{WEN1}) to obtain the elliptic equation satisfied by the electrostatic potential, 
(\ref{3dc1}). Sum (\ref{WEN4}) in $i$ and we find the boundary condition (\ref{3dc2}).
The coefficients $A$ and $B$ in equation (\ref{3dc1}) are now time dependent, but 
we can see from (\ref{WEN1}), that to leading order, the ionic concentrations do 
not change in the membrane potential time scale. Thus, $A$ and $B$ are constant to leading order.
The ease with which one can see the correspondence between the electroneutral model and the 
cable model is an appealing feature of the electroneutral approach.

\subsection{Simplified 3D-Cable Model}

We reach a further simplification by considering the following situation. 
Suppose the long time average of the transmembrane currents $\tilde{j}_i$ is equal to $0$.  
That is to say, if we average over a sufficient long time, 
there is no net current flowing through the membrane. An electrically active cell whose 
ion channel currents are quickly counter-balanced by ionic pumps may fit this 
category. Then, the ionic concentrations should
relax to a stationary value in the slow diffusion time scale.
If there are no fixed charges $\tilde{\rho}$, or if the fixed charges are spatially uniform, the 
resulting ion concentration profile will be spatially uniform within each region.
We apply the above 3-D cable model 
to this situation.  From (\ref{3dc1})-(\ref{3dc3}), we obtain:
\begin{align}
\Delta \Phi^0&=0\label{simp3d1}\\
-A^{(k)}\grad\Phi^0\cdot\mathbf{n}^{(kl)}&=\theta\PD{\Phi^{(kl),0}}{\tau_V}+\alpha I_\text{ion}\label{simp3d2}
\end{align}
The gradient of $B$ vanishes because of the spatial uniformity of $C_i^0$. 
Note that $A^{(k)}$ is a constant that depends only on the region number $(k)$,
and expresses the ohmic conductivity of the electrolyte medium. 
We shall call this the {\em simplified 3-D cable model}.
We note that this system, when homogenized in a quasi-periodic domain, gives rise 
to the bidomain equations, which are widely used in simulations of organ-level 
cardiac electrophysiology \cite{Neu-Krassowska,KS}.

\subsection{Derivation of Standard Cable Model}\label{cbl}
We now derive the traditional cable model by considering the above simplified 3D-cable model under 
specialized geometry. We note of an analysis of a similar situation for a passive cable in which 
a different approach is used to address this issue \cite{Rall}.

Consider an infinitely long cylinder of radius $r^{\text{int}}$. The dimensionless radius will 
therefore be $\eta\equiv r^{\text{int}}/L_0$. 
This infinite cylinder is surrounded by an extracellular space which lies between 
this cylinder and a concentric cylinder of radius $r^{\text{ext}}(>r^{\text{int}})$. 
This extracellular region is insulated at the outer boundary. 
We shall let $\xi=r^{\text{ext}}/r^{\text{int}}$. Equations of the simplified 3D-cable model
(\ref{simp3d1}) and (\ref{simp3d2}) specialized to this situation are:
\begin{align}
\PDD{2}{\Phi}{Z}+\Delta_D\Phi&=0 \text{ in } \Omega^{\text{int}}, \Omega^{\text{ext}}\label{cab1}\\
-A^{(k)}\PD{\Phi}{R}&=\theta\PD{[\Phi]}{\tau_V}+\alpha I_\text{ion} , 
\quad [\Phi]\equiv \Phi^\text{int}-\Phi^\text{ext}\text{ at } R=\eta^\pm \label{cab3}\\
-A^\text{ext}\PD{\Phi}{R}&=0 \text{ at } R=\eta\xi\label{cab4}
\end{align}
To avoid cluttered notation, we have eliminated the superscript $0$. In the above, $R$ is the radial, 
$Z$ the axial coordinate and $\Delta_D$ denotes the Laplacian on the plane $Z=\text{const}$.
Equation (\ref{cab3}) is satisfied at $R=\eta$ 
from both the intracellular ($\eta^{-}$) and extracellular ($\eta^{+}$) sides. The superscript 
$k$ denotes either the intra or extracellular region.

We shall now take $\eta$ to be the small parameter in our system. What follows
is a thin-domain asymptotics calculation used for example in lubrication theory \cite{Howison}. 
We rescale the the radial coordinate to $R=\eta \rho$ in (\ref{cab1})-(\ref{cab4}) so that the cell 
membrane corresponds to $\rho=1$.
\begin{align}
\eta^2\PDD{2}{\Phi}{Z}+\Delta_{\tilde{D}} \Phi&=0 \text{ in } \Omega^\text{int}, \Omega^\text{ext}\label{cab1'}\\
-\frac{A^{(k)}}{\eta}\PD{\Phi}{\rho}&=\theta\PD{[\Phi]}{\tau_V}+\alpha I_\text{ion} , 
\quad [\Phi]\equiv \Phi^\text{int}-\Phi^\text{ext}\text{ at } \rho=1^{\pm} \label{cab3'}\\
-A^\text{ext}\PD{\Phi}{\rho}&=0 \text{ at } \rho=\xi\label{cab4'}
\end{align}
where $\Delta_{\tilde{D}}$ denotes the rescaled Laplacian on $Z=\text{const}$.
We now expand $\Phi$ in powers of $\eta^p$ in the following fashion:
\begin{equation}
\Phi=\Phi^0+\eta^p\Phi^1+\cdots \label{cabexp}
\end{equation}
We let $p=2$ so that we obtain nontrivial expressions when the 
above substituted into (\ref{cab1'}):
\begin{align}
\Delta_{\tilde{D}}\Phi^0&=0\label{cabo0}\\
\PDD{2}{\Phi^0}{Z}+\Delta_{\tilde{D}}\Phi^1&=0\label{cabo1}
\end{align} 
Consider the boundary condition (\ref{cab3'}). Upon substitution of (\ref{cabexp}), we see that 
a distinguished limit can be obtained by taking $\alpha\sim \eta$.  This is in fact, hardly surprising.
In Section \ref{nondimmult}, we introduced $\alpha$ as the volume to surface ratio of the domain of interest. The dimensionless 
radius $\eta$ is exactly equal to this ratio (up to a factor of order 1).  We shall thus take $\eta=\alpha$.
Therefore,
\begin{align}
\PD{\Phi^{0,(k)}}{\rho}&=0 \text{ at } \rho=1^\pm, \rho=\xi \label{cabdo0}\\
-{A^{(k)}}\PD{\Phi^1}{\rho}&=\frac{\theta}{\alpha}\PD{[\Phi^0]}{\tau_V}+ I_{\text{ion}}
\text{ at } \rho=1^\pm, \quad \PD{\Phi^1}{\rho}=0 \text{ at } \rho=\xi \label{cabdo1}
\end{align}
First of all, (\ref{cabo0}) with (\ref{cabdo0}) tells us that $\Phi^0$ is constant for fixed $Z$.
In order to find the $Z$ dependence of $\Phi^0$, we need to look at the next order, (\ref{cabo1}).
The solvability of this equation with respect to $\Phi^1$ requires that the following identities 
between an area and a line integral hold for each $Z=Z_0$.
\begin{align}
\int_{\rho<1, Z=Z_0}\Delta_{\tilde{D}}\Phi^1 dA&= \int_{\rho=1^-, Z=Z_0}\PD{\Phi^1}{\rho} ds\\
\int_{1<\rho<\xi, Z=Z_0}\Delta_{\tilde{D}}\Phi^1 dA&= -\int_{\rho=1^+, Z=Z_0}\PD{\Phi^1}{\rho} ds
+\int_{\rho=\xi, Z=Z_0}\PD{\Phi^1}{\rho} ds
\end{align}
where $dA$ denotes an area integral and $ds$ denotes a line integral.
Applying the above to (\ref{cabo1}) and (\ref{cabdo1}) we find that:
\begin{align}
\pi A^\text{int}\PDD{2}{\Phi^{0,\text{int}}}{Z}
&=2\pi\frac{\theta}{\alpha}\PD{[\Phi^0]}{\tau_V}+\int_0^{2\pi} I_\text{ion}d\psi\\
-\pi(\xi^2-1) A^\text{ext}\PDD{2}{\Phi^{0,\text{ext}}}{Z}
&=2\pi\frac{\theta}{\alpha}\PD{[\Phi^0]}{\tau_V}+\int_0^{2\pi} I_\text{ion}d\psi
\end{align}
Dividing by the prefactors and adding the two expressions, we immediately obtain the cable equations:
\begin{align}
G^\text{eff}\PDD{2}{[\Phi^0]}{Z}&=2\pi \paren{\frac{\theta}{\alpha}\PD{[\Phi^0]}{\tau_V}+I_\text{ion}(Z)}\\
\frac{1}{G^\text{eff}}&=\frac{1}{\pi A^\text{int}}+\frac{1}{\pi (\xi^2-1)A^\text{ext}}, \quad
I_\text{ion}(Z)=\frac{1}{2\pi}\int_0^{2\pi} I_\text{ion} d\psi 
\end{align}
We have thus succeeded in deriving the cable model. We note in particular that $Z$
is measured with respect to the length scale $L_0$, which we can now identify as the 
electrotonic length. The time variable $\tau_V$ is measured with respect to $\beta T_0$
which tells us that $\beta T_0$ is ``diffusion'' time scale for the membrane potential.

If $1 \ll \xi \ll \alpha^{-1} (r^\text{int} \ll r^\text{ext} \ll L_0)$, 
we can take the extracellular space to be an isopotential compartment 
and set $G^{\text{eff}}=\pi A^\text{int}$ without sacrificing the validity of the above cable equations.
In dimensional terms, the above equations take the following familiar form:
\begin{align}
\frac{1}{R}\PDD{2}{\phi_\text{m}}{z}
&=p_\text{m}\paren{C_\text{m}\PD{\phi_\text{m}}{t}+i_\text{ion}},\quad p_\text{m}=2\pi r^\text{int} \\
R&=R^{\text{int}}+R^{\text{ext}}\\
\frac{1}{R^\text{int}}&=S^\text{int}\sum_{i=1}^N 
\frac{(qz_i)^2D_i}{k_BT}c_i^\text{int},\quad S^\text{int}=\pi (r^\text{int})^2\\
\frac{1}{R^\text{ext}}&=S^\text{ext}\sum_{i=1}^N 
\frac{(qz_i)^2D_i}{k_BT}c_i^\text{ext},\quad  S^\text{ext}=\pi ((r^\text{ext})^2-(r^\text{int})^2)
\end{align}
Here, $\phi_\text{m}$ is the membrane potential and $i_\text{ion}$ is the dimensional transmembrane 
current, averaged over the $Z=$const cross-section of the membrane.

We note that the above derivation of the cable model did not assume an axisymmetric solution to the 
equations.  The axisymmetry, or more strongly, the constancy of the electrostatic potential for 
each $Z$ cross-section is a consequence purely of the scaling relations. Related to this is 
the observation that the above can be generalized to arbitrary cross-sectional geometry. All we 
have used is the divergence theorem as applied to each cross section; we have 
made essentially no use of the fact that the cross-section was a disc.

\section{Numerical Validation of Asymptotics}\label{comp}

In this section we shall test the behavior of the electroneutral model against 
that of the Poisson model by way of numerical simulations.
We confine numerical validation to test cases which reduce to one dimensional computations.  
This is because the Poisson model requires extremely small time steps and spatial resolution, 
which makes it computationally overwhelming to compare 
the two models in a full two or three dimensional setting. 
We have considered two geometrical situations, one spherical and one planar, 
but we shall only discuss the spherical case, since results for the planar 
calculations are very similar to the spherical \cite{morithesis}.


We take a spherical cell of radius $l$.
Let the center of the cell be the origin, and let $r$ be the radial coordinate.
We seek solutions to the equations (electroneutral or Poisson) which 
depend only on the radial coordinate $r$. We have thus a one dimensional problem.
The region characterized by $r<l$ is the intracellular space.
We confine our simulation domain to $r<2l$ and impose 
no-flux boundary conditions at $r=2l$.  Thus, our extracellular space 
is the region $l<r<2l$. The $l$ we use here as the radius of the cell 
is to be identified with the $l$ we introduced as the volume to surface ratio
in Section \ref{nondimmult}. 

We now
rescale length so that $L_0$, (\ref{defL0}), is the representative length 
scale. The dimensionless cell radius is now $\alpha=\frac{l}{L_0}$.
We shall continue to use $r$ as our dimensionless coordinate. 
Thus, $r<\alpha$ is the intracellular region and $\alpha<r<2\alpha$
is the extracellular region.
We use the finite volume method
to perform the simulations. 
We subdivide the computational 
region into spherical shells indexed by $k$. The thickness of the spherical 
shells is made to be smaller near the membranes so as to resolve the 
space charge layer and the fast diffusion layer.
The details of the numerical scheme as explained in \cite{morithesis}
will be reported elsewhere.

We consider four ionic species with the following dimensionless diffusion coefficient
and valence.
\begin{align}
\tilde{D}_1&=2, & \tilde{D}_2&=1/2, & \tilde{D}_3&=1, & \tilde{D}_4&=1/2\\
z_1&=1, & z_2&=1, & z_3&=-1, & z_4&=2
\end{align}
Recall from Section \ref{nondimmult} that $\beta, \alpha$ and $\theta^*$ are the dimensionless
parameters that characterize the system of equations. The parameter 
$\theta^*=10^{-2}$(cf (\ref{thetadef}) has a fixed value. We consider 
three pairs of parameter values:
\begin{equation}
(\beta,\alpha)=(10^{-3},10^{-2}), 
\quad (10^{-3.5},10^{-1.5}), 
\quad (10^{-4},10^{-1})
\end{equation}
We expect the electroneutral model to be a good approximation to the Poisson model for small 
values of $\beta$. We thus
take $\beta$ to be slightly larger than the typical values $\beta=10^{-4}\sim 10^{-6}$
to perform a more stringent test of validity of the electroneutral model.



We shall start our simulation at time $\tau_V=-T_r$ where $T_r$ is positive.
The reason for this will become clear shortly.
For the electroneutral model, we set the initial conditions at $\tau_V=-T_r$ for 
$C_i$ to be:
\begin{align}
C_1(r,-T_r)&=1+C_g(2|\alpha-r|-\alpha) & C_2(r,-T_r)&=2-C_1(r,-T_r)\\
C_3(r,-T_r)&=2 & &\\
C_4(r,-T_r)&=10^{-6} \text{ for } r<\alpha & C_4(r,-T_r)&=10^{-3} \text{ for } r>\alpha \\
\tilde{\rho}_0(r)&=-\sum_{i=1}^4z_iC_i(r,-T_r) & &
\end{align}
We let $C_g=0.9$ so that there is a steep initial gradient of the ionic concentrations.
The very small initial values of $C_4$ are motivated by calcium concentration profiles in 
physiological systems.
At the membrane boundary, we must specify $\Phi_\text{m}(\tau_V)$ and $\lambda_i(\alpha\pm,\tau_V)$
at $\tau_V=-T_r$.
\begin{align}
\Phi_\text{m}(-T_r)
&=\Phi(\alpha-)-\Phi(\alpha+)=-\frac{\theta^*}{\theta}\\
\lambda_i(\alpha\pm,-T_r)
&=\frac{z_i^2C_i(\alpha\pm,-T_r)}{\sum_{k=1}^4 z_k^2C_k(\alpha\pm,-T_r)}
\end{align}
where $\alpha+$ and $\alpha-$ denote the $r>\alpha$ and the $r<\alpha$ faces 
of the membrane respectively.

For the Poisson model, we need to specify the initial ionic concentrations.
Given initial conditions for the electroneutral model, 
we set the corresponding initial conditions 
for the Poisson model to be:
\begin{align}
C_i(r,-T_r)&=C_i^\text{electroneutral}(r,-T_r)
-\frac{4\pi\alpha^2\lambda_i(\alpha-,-T_r)\theta^*}{z_i(4\pi/3) \alpha^3} 
\text{ if } r<\alpha\\
C_i(r,-T_r)&=C_i^\text{electroneutral}(r,-T_r)
+\frac{4\pi\alpha^2\lambda_i(\alpha+,-T_r)\theta^*}{z_i(4\pi/3) ((2\alpha)^3-\alpha^3)} 
\text{ if } r>\alpha
\end{align}
The rationale for setting $C_i$ as above is the following.
The initial conditions for the electroneutral model says that each 
ionic species contributes a surface charge amount $\lambda_i\theta^*$
times the membrane area $4\pi\alpha^2$. To set the initial conditions 
for the Poisson model, we need to take into account this contribution.
We spread this surface charge contribution uniformly throughout the 
intracellular and extracellular spaces. 

The problem with this initialization is that the excess charge should 
not be uniformly distributed but should be distributed so that the concentration 
profile shows a space charge layer near the membrane.  Since we do not know the 
exact details of this concentration profile a priori, we let the Poisson system 
relax between $-T_r<\tau_V<0$ to a 
state where the bulk is approximately electroneutral and the excess 
charge accumulates near the membrane. During this period, we  
set the membrane current equal to zero and the dimensionless diffusion 
coefficients to be equal to $\tilde{D}_i=1$. We let $T_r=10\beta$, 
$10$ times the charge relaxation time.

\begin{figure}
\begin{center}
\includegraphics[width=\textwidth]{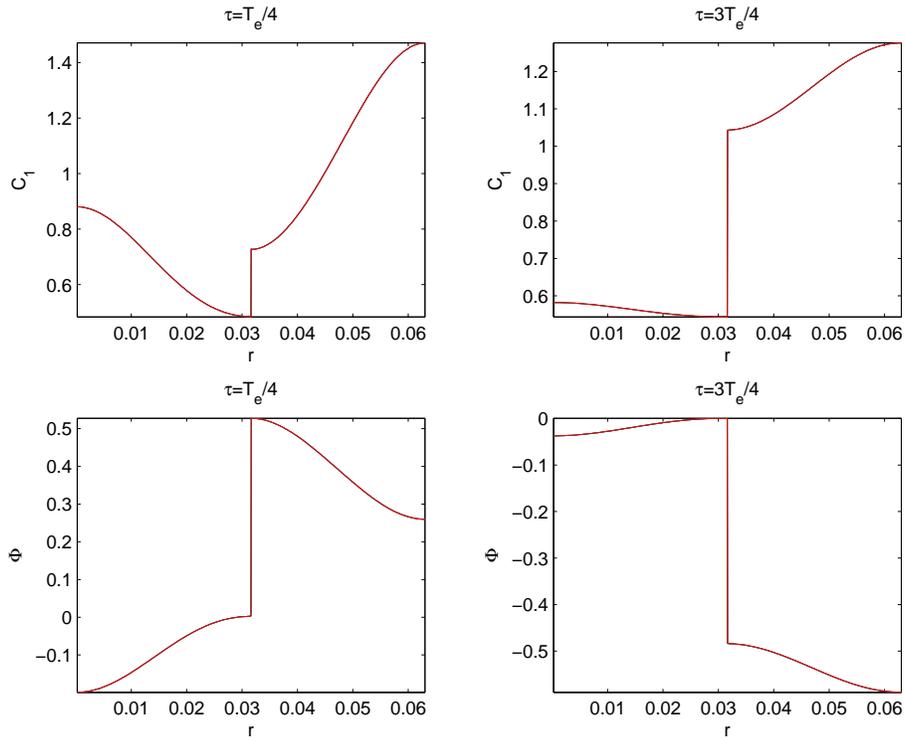}
\end{center}
\caption{Snapshots of simulation when $(\beta,\alpha)=(10^{-3.5},10^{-1.5})$.
Three curves, the Poisson computation, the raw data and modified data from the 
electroneutral models are plotted. The three curves are virtually indistinguishable.
}\label{snapvalidsph}
\end{figure}

At time $t=0$ we turn on a current of constant strength $\alpha$
carried by ionic species $i=4$ 
flowing from the extracellular space ($r>\alpha$) into the intracellular 
space ($r<\alpha$). We let our simulations last 
until $\tau_V=T_e=2\frac{\theta^*}{\alpha}$, which is approximately the time 
it takes to depolarize the dimensionless membrane potential from $-1$ to $1$.
We place $N_r=200$ computational voxels in both the extracellular and intracellular regions
(a total of $2N_r=400$ voxels), 
and we take the time step $\Delta \tau_V=\frac{\beta}{5}$. Using 
a larger time step led to numerical instabilities with the Poisson model.
A snapshot from a sample run of this simulation is shown in Figure \ref{snapvalidsph}.

Raw data produced by the electroneutral model do not 
capture the ionic concentration or electrostatic potential profiles in the Debye 
layer. But it is possible to produce an approximate profile in the Debye layer
based on the asymptotic calculations we performed.
We can see from (\ref{Phiin}) and (\ref{Ciin}) that the Debye layer 
has the effect of adding a correction term to $C_i$ and $\Phi$ that decays 
exponentially with distance from the membrane. The decay length and the 
magnitude of the correction term can be approximated by the 
values of $\Phi, C_i$ evaluated at the membrane and $\lambda_i$. 
For $\Phi$, we modify the raw data from the electroneutral model as follows:
\begin{align}
\Phi^\text{modified}&=\Phi+\delta\Phi\\
\delta\Phi&=-\frac{\theta\Phi_\text{m}}{\Gamma^+}
\exp\paren{-\Gamma^+\frac{|r-\alpha|}{\beta}} \text{ if } r>\alpha\\
&=-\frac{\theta\Phi_\text{m}}{\Gamma^-}
\exp\paren{-\Gamma^-\frac{|r-\alpha|}{\beta}} \text{ if } r<\alpha\\
\Gamma^\pm&=\sqrt{z_i^2C_i(r=\alpha\pm)}
\end{align}
where the double signs correspond in the last line. 
For the ionic concentrations $C_i$, 
\begin{align}
C_i^\text{modified}&=C_i+\delta C_i\\
\delta C_i&=-\frac{\lambda_i(\alpha^+)\theta\Phi_\text{m}\Gamma^+}{z_i}
\exp\paren{-\Gamma^+\frac{|r-\alpha|}{\beta}} \text{ if } r>\alpha\\
&=-\frac{\lambda_i(\alpha^-)\theta\Phi_\text{m}\Gamma^-}{z_i}
\exp\paren{-\Gamma^-\frac{|r-\alpha|}{\beta}} \text{ if } r<\alpha
\end{align}
We note that $\delta C_i$ and $\delta \Phi$ are expressed entirely 
in terms of raw data computed with the electroneutral model.
When comparing the electroneutral model with the Poisson model, we shall use 
the above modified profile.

In order to quantify the difference between the electroneutral and Poisson 
calculations, we introduce the following norm on the computational domain.
Suppose the quantity $u$ is defined at each voxel indexed by $k$.
We define the discrete $p$-norm as: 
\begin{align}
\norm{u}_{L^p}&=\paren{\frac{\sum_{k=1}^{2N_r}|V_k||u_k|^p}
{\sum_{k=1}^{2N_r}|V_k|}}^{1/p}, \quad 1\leq p<\infty \label{valdiscpnorm}\\
\norm{u}_{L^\infty}&=\max_k |u_k|
\end{align}
where $u_k$ is the value of $u$ at the $k$-th voxel and $V_k$ is the volume of the $k$-th voxel.
In defining the $L^p$ norm in (\ref{valdiscpnorm}), we have divided by a normalizing factor
so that $\norm{u}_{L^p}$ gives an average measure of the ``$L^p$ deviation''.
In particular, $\lim_{p\rightarrow \infty} \norm{u}_{L^p}=\norm{u}_{L^\infty}$.
For ionic concentrations $C_i$, we use the relative error:
\begin{align}
\mathcal{E}_p(C_i)=\frac{\norm{C_i^\text{electroneutral}-C_i^\text{Poisson}}_{L^p}}
{\norm{C_i^\text{Poisson}}_{L^p}}
\end{align}
This is a more stringent criteria than using the absolute error (without the 
denominator in the above) especially for $C_4$ whose initial concentration
is very small.

For the electrostatic potential $\Phi$, it does not make sense to use the 
relative error since an arbitrary constant constant may be added to $\Phi$.
We thus, measure the error in $\Phi$ as:
\begin{equation}
\mathcal{E}_p(\Phi)=\min_{c_p\in\mathbb{R}}
\norm{\Phi^\text{electroneutral}-\Phi^\text{Poisson}+c_p}_{L^p}
\end{equation}
Note that it is reasonable to consider the absolute error in $\Phi$, 
since $\Phi$ is dimensionless, and its typical magnitude is $1$.
Though $\mathcal{E}_p(\Phi)$ may in general be difficult to compute 
in closed form, this is possible when $p=1,2,\infty$,  
values of $p$ for which we shall compute $\mathcal{E}_p(\Phi)$ in the following.

\begin{center}
\begin{table}
\begin{tabular}{|c|c||c|c|c|c|c|}
\hline
$(\beta,\alpha)$ & $L^p$ & $\mathcal{M}_p(C_1)$ & $\mathcal{M}_p(C_2)$
& $\mathcal{M}_p(C_3)$ & $\mathcal{M}_p(C_4)$ & $\mathcal{M}_p(\Phi)$ \\ \hline
\multirow{3}{*}{$(\beta_1,\alpha_1)$} 
& $L^1$      
& $3.79\times 10^{-5}$ & $6.98\times 10^{-5}$ & $6.56\times 10^{-5}$ & $1.55\times 10^{-4}$
& $2.73\times 10^{-5}$ \\ \cline{2-7}
& $L^2$      
& $4.67\times 10^{-5}$ & $8.43\times 10^{-5}$ & $7.71\times 10^{-5}$ & $1.74\times 10^{-4}$ 
& $5.95\times 10^{-5}$ \\ \cline{2-7}
& $L^\infty$ 
& $2.87\times 10^{-4}$ & $2.35\times 10^{-4}$ & $2.86\times 10^{-4}$ & $5.85\times 10^{-4}$
& $1.40\times 10^{-4}$ \\ \hline
\multirow{3}{*}{$(\beta_2,\alpha_2)$} 
& $L^1$      
& $7.20\times 10^{-5}$ & $2.76\times 10^{-5}$ & $3.25\times 10^{-5}$ & $2.99\times 10^{-5}$ 
& $4.70\times 10^{-5}$ \\ \cline{2-7}
& $L^2$      
& $8.46\times 10^{-5}$ & $4.40\times 10^{-5}$ & $5.47\times 10^{-5}$ & $5.15\times 10^{-5}$ 
& $1.03\times 10^{-4}$ \\ \cline{2-7}
& $L^\infty$ 
& $2.32\times 10^{-4}$ & $1.57\times 10^{-4}$ & $1.89\times 10^{-4}$ & $2.06\times 10^{-4}$ 
& $1.71\times 10^{-4}$ \\ \hline
\multirow{3}{*}{$(\beta_3,\alpha_3)$} 
& $L^1$      
& $1.74\times 10^{-6}$ & $1.79\times 10^{-6}$ & $4.25\times 10^{-7}$ & $3.31\times 10^{-7}$
& $1.62\times 10^{-5}$ \\ \cline{2-7}
& $L^2$      
& $2.58\times 10^{-6}$ & $2.98\times 10^{-6}$ & $1.97\times 10^{-6}$ & $1.71\times 10^{-6}$
& $4.16\times 10^{-5}$ \\ \cline{2-7}
& $L^\infty$ 
& $1.19\times 10^{-5}$ & $8.59\times 10^{-5}$ & $7.53\times 10^{-5}$ & $1.57\times 10^{-4}$ 
& $6.95\times 10^{-5}$\\ \hline
\end{tabular}
\caption{$\mathcal{M}_p$ values for spherical geometry 
for three computational experiments with different values 
of $\beta$ and $\alpha$. Here, $(\beta_1,\alpha_1)=(10^{-3},10^{-2})$,
$(\beta_2,\alpha_2)=(10^{-3.5},10^{-1.5})$, $(\beta_3,\alpha_3)=(10^{-4},10^{-1})$.
\label{valerrortablesph}}
\end{table}
\end{center}

In table (\ref{valerrortablesph}), we list the $\mathcal{M}_p(u), u=\Phi \text{ or } C_i$
where:
\begin{equation}
\mathcal{M}_p(u)=\max_{0\leq\tau_V\leq T_e} \mathcal{E}_p(u)
\end{equation}
We see that for all parameter ranges tested here, the error falls within order $10^{-4}$. 
This translates to a $0.01\%$ error in $C_i$ and an error of about $0.025 \text{mV}$ in the dimensional 
electrostatic potential $\phi$. 
In cases $(\beta,\alpha)=(10^{-3},10^{-2})$ or $(10^{-3.5},10^{-1.5})$, 
$\alpha$ is comparable in magnitude to $\sqrt{\beta}$.
The degree of correspondence exhibited for these two cases is remarkable since the asymptotic calculations were 
performed under the assumption that $\alpha=\mathcal{O}(1)$ with respect to $\sqrt{\beta}$.
It is notable that the relative error is order $10^{-4}$
even for $C_4$ which has a vanishing small concentration. This tells us that we may 
include ions of very small concentration into our model framework, which is significant 
if we are to include calcium dynamics \cite{Aidley}. 

We see that $\mathcal{M}_\infty(C_i)$ is significantly larger than $\mathcal{M}_1(C_i)$
or $\mathcal{M}_2(C_i)$. Despite the modification we performed on the raw data for the electroneutral model,
the deviation between the electroneutral and Poisson models are still concentrated 
at the Debye layer. 
Since this layer is very small in volume, the $L^1$ and 
$L^2$ errors are not significantly affected. 

\section{Conclusion}\label{disc}

The Poisson model, a candidate model for three dimensional 
cellular electrical activity, is computationally difficult to deal with, because of the 
presence of the Debye layer which develops at membrane interfaces.
We introduced the electroneutral model as a computationally amenable and biophysically natural
model of cellular electrical activity. We use asymptotic calculations to demonstrate 
the validity of the electroneutral model. The matched asymptotic calculations required 
the introduction of two boundary layers at the membrane, the inner Debye layer and the 
intermediate fast diffusion layer. We show that the electroneutral model gives an 
approximation to the Poisson model in the intermediate and outer layers as the small 
parameter $\beta$, the ratio between the Debye length and the electrotonic length, 
becomes small. We demonstrated computationally that the electroneutral model gives an 
excellent approximation to the Poisson model.

\begin{figure}
\begin{center}
\unitlength 0.1in
\begin{picture}( 35.2000, 12.9000)(  2.0000,-16.8000)
%
\special{pn 8}%
\special{pa 200 720}%
\special{pa 840 720}%
\special{pa 840 1200}%
\special{pa 200 1200}%
\special{pa 200 720}%
\special{fp}%
%
\special{pn 8}%
\special{pa 1160 720}%
\special{pa 1800 720}%
\special{pa 1800 1200}%
\special{pa 1160 1200}%
\special{pa 1160 720}%
\special{fp}%
%
\special{pn 8}%
\special{pa 2120 720}%
\special{pa 2760 720}%
\special{pa 2760 1200}%
\special{pa 2120 1200}%
\special{pa 2120 720}%
\special{fp}%
%
\special{pn 8}%
\special{pa 3080 720}%
\special{pa 3720 720}%
\special{pa 3720 1200}%
\special{pa 3080 1200}%
\special{pa 3080 720}%
\special{fp}%
%
\special{pn 8}%
\special{pa 840 960}%
\special{pa 1160 960}%
\special{fp}%
\special{sh 1}%
\special{pa 1160 960}%
\special{pa 1094 940}%
\special{pa 1108 960}%
\special{pa 1094 980}%
\special{pa 1160 960}%
\special{fp}%
%
\special{pn 8}%
\special{pa 1800 960}%
\special{pa 2120 960}%
\special{fp}%
\special{sh 1}%
\special{pa 2120 960}%
\special{pa 2054 940}%
\special{pa 2068 960}%
\special{pa 2054 980}%
\special{pa 2120 960}%
\special{fp}%
%
\special{pn 8}%
\special{pa 2760 960}%
\special{pa 3080 960}%
\special{fp}%
\special{sh 1}%
\special{pa 3080 960}%
\special{pa 3014 940}%
\special{pa 3028 960}%
\special{pa 3014 980}%
\special{pa 3080 960}%
\special{fp}%
\put(3.6000,-10.4000){\makebox(0,0)[lb]{Poisson}}%
\put(12.0000,-9.1000){\makebox(0,0)[lb]{Electro-}}%
\put(21.7000,-10.4000){\makebox(0,0)[lb]{3D-Cable}}%
\put(32.4000,-10.4000){\makebox(0,0)[lb]{Cable}}%
\put(2.0000,-15.2000){\makebox(0,0)[lb]{inner layer}}%
\put(13.2000,-15.2000){\makebox(0,0)[lb]{int. layer}}%
\put(3.6000,-5.6000){\makebox(0,0)[lb]{3D,$c_i,\phi$}}%
\put(13.2000,-5.6000){\makebox(0,0)[lb]{3D,$c_i,\phi$}}%
\put(22.8000,-5.6000){\makebox(0,0)[lb]{3D,$\phi$}}%
\put(32.4000,-5.6000){\makebox(0,0)[lb]{1D,$\phi$}}%
\put(21.0000,-15.2000){\makebox(0,0)[lb]{outer layer}}%
%
\special{pn 8}%
\special{pa 1000 960}%
\special{pa 840 1680}%
\special{fp}%
\special{pa 1960 960}%
\special{pa 840 1680}%
\special{fp}%
\put(7.0000,-18.4000){\makebox(0,0)[lb]{matched asymptotics when $\beta$ small}}%
%
\special{pn 8}%
\special{pa 2920 960}%
\special{pa 2760 1520}%
\special{fp}%
\put(20.0000,-16.9000){\makebox(0,0)[lb]{thin domain asymptotics when $\alpha$ small}}%
\put(12.0000,-10.9000){\makebox(0,0)[lb]{neutral}}%
%
\special{pn 8}%
\special{pa 1000 400}%
\special{pa 1000 1510}%
\special{fp}%
%
\special{pn 8}%
\special{pa 1960 400}%
\special{pa 1960 1510}%
\special{fp}%
%
\special{pn 8}%
\special{pa 2920 400}%
\special{pa 2920 1510}%
\special{fp}%
\end{picture}%
\end{center}
\caption{Hierarchy of Electrophysiology Models}\label{Fighierarchy}
\end{figure}
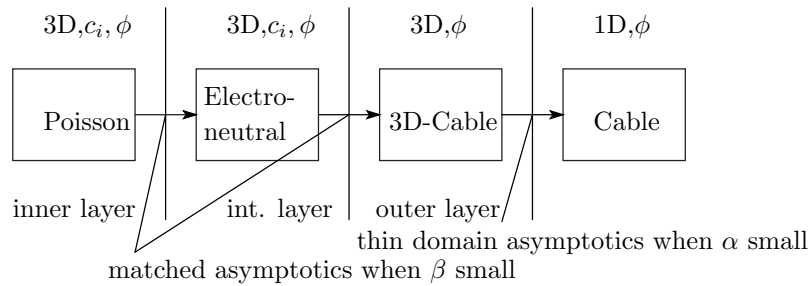

We have also succeeded in systematically deriving the 
standard cable model from the Poisson model or the electroneutral model.  The above derivation can be viewed as a
significant step toward a full study of the validity of the cable model,  
an issue of fundamental importance to computational neuroscience \cite{Scott}.  
In the course of this derivation, we have seen that there are models of intermediate complexity in between the 
Poisson or electroneutral model and the cable model (Figure \ref{Fighierarchy}).  The 3D-cable model and 
the simplified 3D-cable model describe the dynamics of the electrostatic potential in a three dimensional 
setting, but ignores the dynamics of ionic concentrations. 
We believe that each of these models will be suitable in certain situations, 
the Poisson or electroneutral models being the most detailed.

When $l_m\sim \sqrt{\beta}$ matching at the inner-intermediate layer interface 
resulted in an additional surface drift-diffusion term 
along the membrane (cf. \ref{matsmall}). It would be interesting to incorporate this into the electroneutral model
and see whether this term leads to a significant difference in the behavior of the model.
When $l_m\sim \sqrt{\beta}$, matching between the intermediate and outer layers is probably challenging, 
since ionic fluxes parallel to the membrane will be comparable in magnitude to fluxes 
perpendicular to the membrane. The intermediate layer will lose its one-dimensional structure.
We believe that the electroneutral model correctly captures 
the dynamics of ionic concentrations in the slow diffusion time scale (time scale $T_0$). 
This claim is supported in part by the fact that 
the conservation relation, equation (\ref{consvion}) is satisfied. We plan to
investigate these points in future work.  

\section{Appendix}\label{app1}
The calculations presented below are identical to the one that appears in \cite{MPJ1},
except for notational differences and some additions.
We would like to solve (\ref{in2}), (\ref{in4}) under the boundary conditions (\ref{in5})-(\ref{mat2}).
Since this is a one dimensional boundary value problem, we shall think of $\Phi^{b0}$ and $C_i^{b0}$
as functions only of $\xi_1^b$ and do not explicitly write their dependence on $\xi_2^b$ or $\xi_3^b$.

Equation (\ref{in2}) can be integrated easily to obtain
\begin{equation}
C_i^{b0}(\xi_1^b)=C_i^{b0}(\infty)\exp\paren{-z_i(\Phi^{b0}(\xi_1^b)-\Phi^{b0}(\infty))}.
\end{equation}
This equation can be substituted into (\ref{in4}) to yield:
\begin{equation}
-\PDD{2}{\Phi^{b0}}{\xi_1^b}
=\paren{\tilde{\rho}_0+\sum_{i=1}^{N}z_iC_i^{b0}(\infty)\exp\paren{-z_i(\Phi^{b0}(\xi_1^b)-\Phi^{b0}(\infty))}}.
\label{poissonboltzmann}
\end{equation}
Here we use an approximation to linearize the above Poisson-Boltzmann equation. We suppose
\begin{equation}
\left|z_i(\Phi^{b0}(\xi_1^b)-\Phi^{b0}(\infty)) \right| \ll 1 \label{phidev}.
\end{equation}
This can be justified if $\theta^*$ is small, as was shown in \cite{MPJ1}.
The smallness of $\theta^*$ states that the amount of charge that may accumulate 
at the membrane is small. The smallness of this charge accumulation guarantees that 
the deviation of $\Phi$ in the inner layer from its value in the intermediate layer is small.  
Given that (\ref{phidev}) is a valid assumption, we linearize (\ref{poissonboltzmann}) to find:
\begin{align} 
C_i^{b0}(\xi_1^b)&=C_i^{b0}(\infty)\paren{1-z_i(\Phi^{b0}(\xi_1^b)-\Phi^{b0}(\infty))}\label{linear_c_i}\\
\PDD{2}{}{\xi_1^b}(\Phi^{b0}(\xi_1^b)-\Phi^{b0}(\infty))&=\Gamma^2(\Phi^{b0}(\xi_1^b)-\Phi^{b0}(\infty))\label{linphieq}\\
\Gamma^2=\sum^{N}_{i=1}z_i^2C_i^{b0}(\infty)&=\sum^{N}_{i=1}z_i^2C_i^{a0}(0), \quad \Gamma>0 \label{gammaapp}.
\end{align}
Here $C_i^{a0}(0)$ is shorthand for $C_i^{a0}(\xi_1^a=0)$.
To derive (\ref{linear_c_i}) and (\ref{linphieq}), we have used
\begin{equation}
\tilde{\rho}_0+\sum_{i=1}^N z_iC_i^{b0}(\infty)=\tilde{\rho}_0+\sum_{i=1}^N z_iC_i^{a0}(0)=0
\end{equation}
which follows as a consequence of (\ref{ndma3}) and the matching condition (\ref{mat1}).
Solving (\ref{linphieq}) with (\ref{in4}) and (\ref{mat2}), 
\begin{equation}
\Phi^{b0}(\xi_1^b)=\Phi^{a0}(0)-\frac{\tilde{\sigma}}{\Gamma}\exp(-\Gamma \xi_1^b)\label{Phiinapp}
\end{equation}
where $\Phi^{a0}(0)$ is shorthand for $\Phi^{a0}(\xi_1^a=0)$ and $\tilde{\sigma}$ is equal to
\begin{equation}
\tilde{\sigma}=\theta^* (\Phi^{b0}(0)-\Phi^{(l),b0}(0)). \label{mem1}
\end{equation}
Hence, according to (\ref{linear_c_i}) and the matching condition (\ref{mat1}), 
\begin{equation}
C_i^{b0}(\xi_1^b)=C_i^{a0}(0)\paren{1+\frac{z_i\tilde{\sigma}}{\Gamma}\exp(-\Gamma \xi_1^b)}\label{Ciinapp}
\end{equation}
We note that $\tilde{\sigma}$ is the total excess charge found in the inner layer, as 
can be seen as follows.
The excess charge contributed by the $i$-th species of ion $\tilde{\sigma}_i$ can be 
computed using expression (\ref{Ciinapp}) as:
\begin{equation}
\tilde{\sigma}_i\equiv\int_0^\infty z_i(C_i^{b0}(\xi_1^b)-C_i^{a0}(0))d\xi_1^b
=\frac{z_i^2C_i^{a0}(0)}{\Gamma^2}\tilde{\sigma}\equiv \tilde{\lambda}_i\tilde{\sigma}
\end{equation}
From (\ref{gamma}), we see that $\tilde{\lambda}_i$ is given by:
\begin{equation}
\tilde{\lambda}_i=\frac{z_i^2C_i^{a0}(0)}{\Gamma^2}=\frac{z_i^2C_i^{a0}(0)}{\sum_{i'=1}^{N} z_{i'}^2C_{i'}^{a0}(0)}
\end{equation}
We immediately conclude that $\sum_{i=1}^N \tilde{\lambda}_i=1$.
The total excess charge is given by summing $\tilde{\sigma}_i$ in $i$.
\begin{equation}
\sum_{i=1}^N \tilde{\sigma}_i=\paren{\sum_{i=1}^N \tilde{\lambda}_i}\tilde{\sigma}=\tilde{\sigma}.
\end{equation}
The factor $\tilde{\lambda}_i$ thus represents the fraction of excess charge contributed by the $i$-th species of ion.

We now have the solutions $C_i^{b0}$ and $\Phi^{b0}$ except that $\tilde{\sigma}$ is expressed 
in terms of $\Phi^{b0}$. We shall now express $\tilde{\sigma}$ in terms of $C_i^{a0}(0)$ and $\Phi^{a0}(0)$.
First, we observe by substituting $\xi_1^b=0$ in (\ref{Phiinapp}) that
\begin{equation}
\tilde{\sigma}=\Gamma(\Phi^{a0}(0)-\Phi^{b0}(0)). \label{mem2}
\end{equation}
We next rewrite 
$\Phi^{a0}(0)=\Phi^{(k),a0}(0), \Phi^{b0}(0)=\Phi^{(k),b0}(0), \tilde{\sigma}=\tilde{\sigma}^{(k)},
\Gamma=\Gamma^{(k)}$ 
and consider (\ref{mem1}) and (\ref{mem2}) as well as the corresponding expressions 
on the other side of the membrane(the $\Omega^{(l)}$ side).
\begin{align}
\tilde{\sigma}^{(k)}&=\theta^* (\Phi^{(k),b0}(0)-\Phi^{(l),b0}(0))\\
\tilde{\sigma}^{(k)}&=\Gamma^{(k)} (\Phi^{(k),a0}(0)-\Phi^{(k),b0}(0))\\
\tilde{\sigma}^{(l)}&=\theta^* (\Phi^{(l),b0}(0)-\Phi^{(k),b0}(0))\\
\tilde{\sigma}^{(l)}&=\Gamma^{(l)} (\Phi^{(l),a0}(0)-\Phi^{(l),b0}(0))
\end{align}
After some algebra, we find,
\begin{align}
\tilde{\sigma}^{(k)}&=-\tilde{\sigma}^{(l)}=\theta(\Phi^{(k),a0}(0)-\Phi^{(l),a0}(0))\label{cap1}\\
\frac{1}{\theta}&=\frac{1}{\theta^*}+\frac{1}{\Gamma^{(k)}}+\frac{1}{\Gamma^{(l)}}.\label{cap2}
\end{align}
The meaning of relation (\ref{cap2}) becomes clear once this is written in dimensional terms:
\begin{equation}
\frac{1}{C_\text{m}}=\frac{1}{C_\text{m}^*}+\frac{1}{\epsilon\gamma^{(k)}}+\frac{1}{\epsilon\gamma^{(l)}}.
\end{equation}
where $\Gamma=\gamma r_d$.  This relation states that the effective membrane capacitance $C_\text{m}$ 
can be computed as the intrinsic membrane capacitance $C_\text{m}^*$ and the capacitance of the space charge 
layers $\epsilon\gamma^{(k)}, \epsilon\gamma^{(l)}$ in series. We note that in (\ref{cap2}), 
$\theta^*$ is small in magnitude whereas $\Gamma^{(k)}$ and $\Gamma^{(l)}$ are order $1$.
Therefore, $\theta \approx \theta^*$, and $C_\text{m}\approx C_\text{m}^*$.

\section{}\label{app2}
In \cite{MPJ1}, $\tilde{\lambda}_i$ was used in place of $\lambda_i$ in (\ref{deflambdai1}).
This expression substituted into the boundary condition (\ref{WEN4}) yields:
\begin{equation}
z_i\mathbf{F}_{i}\cdot \mathbf{n}^{(kl)}=\theta\PD{\tilde{\lambda}_i\Phi^{(kl)}}{\tau_V}+\alpha\tilde{j}_i
\label{bc}
\end{equation}
By following the same procedure as in Section \ref{ENPo},
it can be easily seen that (\ref{WEN1})-(\ref{WEN3}) together with (\ref{bc})
has the desired approximation properties.
Unfortunately, this system is ill-posed. 

We shall exhibit the ill-posed behavior in a simple situation.
Assume we have two regions, one intracellular and one extracellular.
Let there be no transmembrane currents.
Suppose that there are two positive ionic species with
identical physical properties: the valence and diffusion coefficient 
are equal and scaled to $1$.
Assume moreover that 
the positive ionic charges are counterbalanced completely 
by a spatially uniform immobile charge of magnitude $-1$. 
Equations (\ref{WEN1})-(\ref{WEN3}) and (\ref{bc}) become:
\begin{align}
0&=\PD{C_i}{\tau_V}+\beta\nabla_{\mathbf{X}} \cdot \mathbf{F}_i \label{ill1}\\
\mathbf{F}_i&=-\paren{\nabla_{\mathbf{X}} C_i +C_i\nabla_{\mathbf{X}} \phi} \label{ill2}\\
1&=C_1+C_2 \label{ill3}
\end{align}
Let $\mathbf{n}$ be the outward normal pointing from intracellular to extracellular 
and the membrane potential $[\Phi]=\Phi^\text{i}-\Phi^\text{e}$.
The boundary conditions on the intracellular and extracellular sides of the membrane are
respectively:
\begin{align}
\PD{\tilde{\sigma}_i^{(\text{i})}}{\tau_V}&=\mathbf{F}_i^{\text{i}}\cdot \mathbf{n} &
\tilde{\sigma}^{(\text{i})}_i&=C^{\text{i}}_i\theta[\Phi] \label{illbi1} \\
\PD{\tilde{\sigma}_i^{(\text{e})}}{\tau_V}&=-\mathbf{F}_i^{\text{e}}\cdot \mathbf{n} &
\tilde{\sigma}^{(\text{e})}_i&=-C^{\text{e}}_i\theta[\Phi] \label{illbe1}
\end{align}
We solve the above with the following initial condition:
\begin{align}
[\Phi](\mathbf{X},0)&=\Phi_0\neq 0 \label{init1}\\
C_i(\mathbf{X},0)&=C_{i,0}(\mathbf{X}), C_{1,0}(\mathbf{X})+C_{2,0}(\mathbf{X})=1
\end{align}
We thus assume that the membrane potential is initially constant($=\Phi_0$) throughout, whereas 
the ionic concentration may be nonuniform. From a physical standpoint, the system 
should relax to an equilibrium state in which the ionic concentration gradients 
have disappeared.  

We now show that this initial value problem is ill-posed. 
Summing equation (\ref{ill1}) in $i$ and using (\ref{ill3}) one immediately concludes: 
\begin{equation}
\Delta_{\mathbf{X}} \Phi=0 \label{illphi}
\end{equation}
To obtain boundary conditions for the above Laplace equation, we take 
the summation of both (\ref{illbi1}) and (\ref{illbe1}) in $i$ to obtain: 
\begin{equation}
\theta\PD{[\Phi]}{\tau_V}=-\PD{\Phi^\text{i}}{\mathbf{n}}
= \PD{\Phi^\text{e}}{\mathbf{n}}\label{illphib}
\end{equation}
The equations (\ref{illphi}), (\ref{illphib}), (\ref{init1}) together form an 
initial value problem for $\Phi$ and this has a unique solution: $\Phi$ does not change, 
and is constant within each spatial region.

We now turn to $C_i$. From equation (\ref{ill1}) we obtain:
\begin{equation}
\PD{C}{\tau_V}=\beta \lap_{\mathbf{X}} C, \quad C=C_1, \quad C_2=1-C_1 \label{illc1}
\end{equation}
where we used $\Phi=\text{const}$ within each region. The boundary conditions are:
\begin{align}
\theta\Phi_0\PD{C^{\text{i}}}{\tau_V}&=-\PD{C^{\text{i}}}{\mathbf{n}}\label{illciint}\\
\theta\Phi_0\PD{C^{\text{e}}}{\tau_V}&=-\PD{C^{\text{e}}}{\mathbf{n}}\label{illciext}
=\PD{C^{\text{e}}_1}{\mathbf{n^\text{e}}}
\end{align}
where $\mathbf{n^\text{e}}=-\mathbf{n}$ is the normal pointing from the extracellular to intracellular space.
The evolution equations for the concentrations completely decouple into two separate 
diffusion problems for which the boundary conditions have the form  
$k\PD{C}{t}+\PD{C}{\mathbf{n}}=0$.  When $k$ is negative, this
problem is ill-posed, as was formally established recently in \cite{JLVazquez}.
We see from (\ref{illciint}) and (\ref{illciext}) that one of the diffusion problems 
is bound to be ill-posed unless $\Phi_0=0$ identically.
Here, we shall illustrate this by way of a simple example.  
 
Consider the above in $\mathbf{X}=(X,Y)\in \mathbb{R}^2$ and let the upper and lower half planes
correspond to the intracellular and extracellular spaces respectively. We let
$\Phi_0=-1$, and seek solutions to (\ref{illc1}) and (\ref{illciint}) in the upper half plane
subject to the condition that $C$ decays to $0$ as $Y \rightarrow \infty$.  
We obtain a family of solutions parametrized by $l>1$:
\begin{equation}
C_l(\mathbf{X},\tau_V)
=\exp\paren{\frac{l}{\beta \theta^2}\tau_V
-\frac{l}{\beta\theta}Y}\sin\paren{\frac{X}{\beta\theta}\sqrt{l^2-l}} \label{illsol}
\end{equation}
If the initial data contain any non-zero frequency component along the membrane,
this component will grow exponentially, the exponent being roughly proportional to the wave number.
Thus, the problem is ill-posed.

This instability is most probably a generic feature of the equations 
not confined to the simple situation above.
The instability is caused by the $\PD{C}{\tau_V}$ term in the boundary conditions,
which came from the $\PD{\tilde{\lambda}_i}{\tau_V}$ term.
In general, the boundary conditions are complicated functions of the ionic concentrations, 
but the leading order terms $\PD{C}{\tau_V}$ and $\PD{C}{\mathbf{n}}$ will dominate in 
stability considerations. 
Since the membrane potential $[\Phi]$ is multiplying 
the $\PD{\tilde{\lambda}_i}{\tau_V}$ term, the diffusion problem is 
bound to be ill-posed at least on one side of the membrane.

We now take a closer look at the above situation in an attempt to obtain a well-posed system of equations.
Equation (\ref{illsol}) tells us that the time constant associated with exponential growth in 
the ill-posed solution is at most $\beta\theta^2$, since $l>1$.  
This time duration belongs to the charge relaxation regime 
(actually even faster, by a factor of $\theta^2$).
The spatial scale that appears in (\ref{illsol}) is on the order
of the Debye length or shorter.  The instabilities that 
develop are thus inconsistent with our ansatz that the 
evolution of $C_i$ and $\Phi$ do not possess 
spatiotemporal scales associated with charge relaxation in the space charge layer.

We would like to remove the explosive behavior caused by 
$\PD{\tilde{\lambda}_i}{\tau_V}$. We propose the following fix. Let $\lambda_i$ be 
a quantity that evolves according to the following differential equation.
\begin{equation}
\PD{\lambda_i}{\tau_V}=\frac{\tilde{\lambda}_i-\lambda_i}{\tau_\lambda},\quad \tau_\lambda=\beta, \quad 
\quad \tilde{\lambda}_i=\frac{z_i^2C_i}{\sum_{i'=1}^{N} z_{i'}^2C_{i'}}
\label{eqlam}
\end{equation}
Thus $\lambda_i$ tracks $\tilde{\lambda}_i$ with a time lag $\tau_\lambda=\beta$, 
the charge relaxation time. 
This has the effect of filtering out any temporal structure that exists on 
a time scale smaller than $\mathcal{O}(\beta)$. 
Instead of $\tilde{\lambda}_i$, we shall use $\lambda_i$ in (\ref{bc}).
We note that since the relaxation time constant ($=\beta$) is taken equal for all 
ionic species, the important relation $\sum_i \lambda_i=1$ holds true 
as long as this relation is satisfied at the initial time (see equation (\ref{lambdasum1})).  

It is important to demonstrate that this replacement does not 
change the formal approximation properties of the original system of equations.
We can find the discrepancy between $\lambda_i$ and $\tilde{\lambda}_i$ as follows.
We can solve (\ref{eqlam}) so that:
\begin{equation}
\lambda_i(\tau_V)=\lambda_i(0)\exp\paren{-\tau_V/\beta}
+\frac{1}{\beta}\int_0^{\tau_V} \tilde{\lambda}_i(s)\exp\paren{\frac{s-\tau_V}{\beta}}ds
\end{equation}
If $\tau_V$ is order $1$, expanding $\lambda_i(s)$ around $\tau_V$, one can easily see that:
\begin{equation}
\begin{split}
\lambda_i(\tau_V)=&\tilde{\lambda}_i(0)\exp\paren{-\tau_V/\beta}
+\tilde{\lambda}_i(\tau_V)\paren{1-\exp\paren{-\tau_V/\beta}}\\
&+\beta\PD{\tilde{\lambda}_i}{\tau_V}(\tau_V)\paren{1-\exp\paren{-\tau_V/\beta}}+\cdots
\end{split}
\end{equation}
We see that
\begin{equation} 
\lambda_i(\tau_V)=\tilde{\lambda}_i(\tau_V)+\mathcal{O}(\beta)
\end{equation}
as long as 
$\PD{\tilde{\lambda}_i}{\tau_V}$ is $\mathcal{O}(1)$.
Likewise, 
\begin{equation}
\begin{split}
\PD{\lambda_i}{\tau_V}(\tau_V)=&-\frac{1}{\beta}\tilde{\lambda}_i(0)\exp\paren{-\tau_V/\beta}
+\PD{\tilde{\lambda}_i}{\tau_V}\paren{1-\exp\paren{-\tau_V/\beta}}\\
&+\beta\PDD{2}{\tilde{\lambda}_i}{\tau_V}(\tau_V)\paren{1-\exp\paren{-\tau_V/\beta}}+\cdots
\end{split}
\end{equation}
from which we find that
\begin{equation}
\PD{\lambda_i}{\tau_V}(\tau_V)=\PD{\tilde{\lambda}_i}{\tau_V}(\tau_V)+\mathcal{O}(\beta)\label{diffdeltlambda}
\end{equation}
as long as 
$\PDD{2}{\tilde{\lambda}_i}{\tau_V}$ is $\mathcal{O}(1)$. It is also possible to show that the error 
is $\mathcal{O}(\beta)$ when $\tau_V=\mathcal{O}(\beta)$ provided $\lambda_i(0)=\tilde{\lambda}_i(0)$.
Since $\lambda_i$ and $\PD{\lambda_i}{\tau_V}$ follow $\tilde{\lambda}_i$ and $\PD{\tilde{\lambda}_i}{\tau_V}$ to order 
$\mathcal{O}(\beta)$, replacing (\ref{bc}) with (\ref{eqlam}) will not alter the formal approximation 
properties of the ill-posed model.

Now we perform the same half plane analysis for the model we just proposed 
as was done for the ill-posed system. 
We take $\tau_\lambda$ in (\ref{eqlam}) as a parameter for now, and 
see what values of $\tau_\lambda$ will remove the instability.  
The expression corresponding to (\ref{illsol}) is:
\begin{align}
C_l(\mathbf{X}_V,\tau_V)&=\exp\paren{l\tau_V-mY^a}\sin(kX^a)\\
k^2&=m^2-l,\quad m=\frac{\sqrt{\beta}\theta l}{\tau_\lambda l+1}
\end{align}
Exponential growth corresponds to $l>0$.
We would therefore like to make sure that the following equation for $l$ does not have a positive solution
for any real $k$.
\begin{equation}
\paren{\frac{\sqrt{\beta}\theta l}{\tau_\lambda l+1}}^2-l=k^2
\end{equation}
This is equivalent to showing that the left hand side of the above is non positive when $l\geq 0$.
Note that $l\geq 0$ implies:
\begin{equation}
\paren{\frac{\sqrt{\beta}\theta l}{\tau_\lambda l+1}}^2-l
=\frac{\beta\theta^2l^2}{(\tau_\lambda l+1)^2}-l
\leq\paren{\frac{\beta \theta^2}{4\tau_\lambda}-1}l
\end{equation}
Therefore, $\tau_\lambda=\beta$ is more than adequate to make the above expression non positive,
since $\theta$ is a small number much less than $1$.
We thus see that for the above situation in which model the system 
(\ref{WEN1})-(\ref{WEN3}), (\ref{bc}) fails,
the new model is stable. 

What we have done is to add a stabilizing term to 
an asymptotically correct but ill-posed system. 
The situation here is analogous to having a consistent but unstable numerical discretization 
for an evolution equation. In such cases, one often adds to the numerical scheme
a stabilizing term (e.g. small diffusive correction)
whose order is small so that it does not 
alter the consistency of the scheme \cite{Leveque}.

\bibliographystyle{plain}
\bibliography{thesis}
\end{document}